\documentclass[onecolumn,aps,nofootinbib]{revtex4-2}
\usepackage{graphicx}
\usepackage{epstopdf}
\usepackage{hyperref}
\usepackage{latexsym}
\usepackage{amsmath}
\usepackage{amssymb}

\usepackage[utf8]{inputenc}
\hypersetup{
    colorlinks=true,
    linkcolor=red,
    citecolor=blue,
}
\usepackage{color}
\usepackage[T1]{fontenc}
\usepackage{txfonts}
\usepackage{supertabular}
\usepackage{setspace}
\usepackage{multirow}
\usepackage{makecell}
\usepackage{mathrsfs}

\usepackage[center]{subfigure}

 \newcommand{\bq}{\begin{equation}}
 \newcommand{\eq}{\end{equation}}
 \newcommand{\bqn}{\begin{eqnarray}}
 \newcommand{\eqn}{\end{eqnarray}}

\makeatletter

\newcommand{\Rmnum}[1]{\expandafter\@slowromancap\romannumeral #1@}
\newcommand{\cev}[1]{\reflectbox{\ensuremath{\vec{\reflectbox{\ensuremath{#1}}}}}}
\makeatother

\begin{document}
\title{Geometric approach for the modified second generation time delay interferometry}
\author{Pan-Pan Wang\textsuperscript{1}}
\author{Wei-Liang Qian\textsuperscript{2,3,4}}\email[E-mail: ]{wlqian@usp.br}
\author{Yu-Jie Tan\textsuperscript{1}}
\author{Han-Zhong Wu\textsuperscript{1}}
\author{Cheng-Gang Shao\textsuperscript{1}}\email[E-mail: ]{cgshao@hust.edu.cn}

\affiliation{$^{1}$ MOE Key Laboratory of Fundamental Physical Quantities Measurement, Hubei Key Laboratory of Gravitation and Quantum Physics, PGMF, and School of Physics, Huazhong University of Science and Technology, Wuhan 430074,  P. R. China}
\affiliation{$^{2}$ Escola de Engenharia de Lorena, Universidade de S\~ao Paulo, 12602-810, Lorena, SP, Brazil}
\affiliation{$^{3}$ Faculdade de Engenharia de Guaratinguet\'a, Universidade Estadual Paulista, 12516-410, Guaratinguet\'a, SP, Brazil}
\affiliation{$^{4}$ Center for Gravitation and Cosmology, College of Physical Science and Technology, Yangzhou University, Yangzhou 225009, China}

\date{May 16th, 2022}

\begin{abstract}
The time delay interferometry (TDI) is an algorithm proposed to suppress the laser frequency noise in space-borne gravitational wave detectors.
As a post-processing technique, it is implemented by constructing a virtual equal arm interferometer through an appropriate combination of the time-shifted data streams.
Such an approach is tailored to the intrinsic feature of the space-based gravitational wave detection, namely, the distances between spacecraft are governed by the orbit dynamics and thus can not be held constant.
Among different implementations, the geometric TDI was introduced as a method of exhaustion to evaluate the second-generation TDI combinations.
The applications of the algebraic approach based on computational algebraic geometry, on the other hand, are mostly restricted to the first and the modified first-generation TDI.
Besides, geometric TDI furnishes an intuitive physical interpretation about the synthesis of the virtual optical paths.
In this paper, the geometric TDI is utilized to investigate the modified second-generation TDI combinations in conjunction with a ternary search algorithm.
The distinction between the second-generation and modified second-generation TDI solutions is elaborated regarding the rate of change of the arm lengths with respect to the opposite cyclic directions.
For the sixteen-link combinations, forty second-generation TDI solutions are recovered, among which nine of them are identified as the modified second-generation ones.
Furthermore, we explore the properties of the modified second-generation TDI solutions, which turn out to be potentially preferable in practice.
Regarding the Taylor expansion of arm lengths in time, the expressions for the leading-order optical path residuals for the relevant geometric TDI combinations are derived, which is further specified using the Keplerian orbits of the spacecraft for the LISA detector constellation.
The response function, noise power spectral density, and signal-to-noise ratio of the TDI solutions are given analytically and discussed.
One obtains three distinct sensitivity curves among nine sixteen-link modified second-generation TDI combinations, while eight sensitivity curves are encountered out of thirty-one second-generation ones.
It is argued that the modified second-generation TDI solutions present a quantitative advantage over their second-generation counterparts. 
Even though the noise suppressions of both scenarios are found to be at the same level, owing to the cancellations in the response function caused by the temporal symmetry of the arm lengths, the magnitude of the gravitational wave signals is less pronounced for the second-generation TDI solutions.
Moreover, analytic analysis confirms that the alternative modified second-generation TDI solutions are desirable as they possess fewer zeros in the average response function and the noise power spectral density, in accordance with previous findings.

\end{abstract}

\maketitle


\section{Introduction}\label{section1}
Since 2015, a series of gravitational wave (GW) detections~\cite{LIGO-01, LIGO-02, LIGO-03,LIGO-04,LIGO-05,LIGO-06,LIGO-07} performed by the Advanced LIGO and Advanced Virgo collaboration has opened up a new era of observational astronomy.
These high-frequency GWs are originated from dramatic processes such as the merger of black holes and neutron stars.
On the other hand, ordinary astrophysical events usually give rise to GW radiations at a lower frequency band.
As a result, the successful observations by the ground-based detectors have further incentivized various projects aiming at space-borne facilities.
Unlike the ground-based GW facilities, whose typical observation frequencies are $10\rm{Hz}-\rm{kHz}$ , the target frequency range of the space-based detectors is between 0.1 mHz and 10 Hz. 
Several projects are under active development, which include LISA~\cite{gw-lisa1,gw-lisa2}, TianQin~\cite{gw-tianqin}, Taiji~\cite{gw-Taiji}, and DECIGO~\cite{gw-DECIGO}.

For the GWs associated with the $0.1\rm{mHz-kHz}$ frequency band, direct observation can be accomplished by using laser interferometers.
The existing implementation of space-based detection involves the measurements of Doppler frequency shifts via the interferences among six laser beams exchanged between three spacecraft.
An individual spacecraft follows a trajectory determined by the geodesic. Subsequently, the arm lengths vary in time so that an ideal equal-arm Michelson interferometer setup is not feasible in space.
Therefore, the dominant noise source, characteristic of the space-borne interferometers, comes from the laser frequency noise.
The noise is embedded in the interferometric signals between the distant and local laser beams, known as the science data stream, and it is typically many orders of magnitude above the GW signal~\cite{tdi-02}.
In this regard, the TDI algorithm~\cite{tdi-01,tdi-02,tdi-03} is a scheme to efficiently suppress the laser frequency noise by constructing virtual equal arm interference through the appropriate combination of the data streams.
As an offline post-processing technique, it is tailored to deal with the scenarios of unequal armlengths. 

The first generation TDI combinations is applicable to the case of static detector constellation with constant unequal armlength~\cite{tdi-01,frame-01-2000,tdi-d55-2001,res-semi--01-2002,tdi-geome-2002,tdi-d22,tdi-laser-06,tdi-laser-LISACode}.
The modified first-generation TDI discriminates between opposite optical paths along a given arm length, and therefore, it is suitable to handle a detector layout that rigidly rotates at a constant speed~\cite{tdi-geometric-2005}.
Both the first and the modified first-generation TDI can be treated rigorously using the computational algebraic through the notation of the first module of syzygies~\cite{tdi-geome-2002, tdi-laser-04}.
Alternatively, one may employ the geometric TDI approach~\cite{tdi-geometric-2005} first proposed by Vallisneri.
As a method of exhaustion, it provides an intuitive interpretation that the resulting TDI combination effectively synthesizes an equal arm interferometer which cancels out the laser frequency noise.
In practice, the spacecraft are in motion, and therefore, the two synthetic optical paths furnished by the first generation TDI are not entirely identical.
These discrepancies are originated from the nonvanishing commutators of time-delay operators of different armlengths.
In this regard, the second generation TDI is aiming at a constellation with rotating and flexing armlengths~\cite{tdi-laser-01,tdi-d99,tdi-d88,tdi-2010-Dhurandhar,tdi-clock4,tdi-filter-s4}.
Specifically, compared with the first generation TDI, it further considers the discrepancies due to the rate of change of the distance between the spacecraft.
However, due to the non-commutative nature of the time-delay operators, the commutative ring theory cannot be straightforwardly applied to the second-generation TDI.
On the other hand, the geometric TDI can be readily utilized to seek feasible solutions by explicitly requiring that the additional contributions to the optical paths be canceled out up to the first order in the velocity.
The topic of geometric TDI has been explored recently by several authors. 
In Ref.~\cite{tdi-geometric-2020}, Muratore {\it et al.} enumerated possible TDI combinations using symbolic algorithms and investigated the residual delay numerically.
The approach was further extended recently~\cite{tdi-geometric-2022a} aiming primarily at null-combinations that carry mostly the information on instrumental noise while the GWs are suppressed. 
In Ref.~\cite{tdi-geometric-2022b}, Hartwig and Muratore characterized geometric TDI channels in terms of the first generation TDI variables.

Similar to the generalization from the first to the modified first-generation TDI, one may distinguish the rate of change of the arm length in the two opposite cyclic directions for a given arm length.
By explicitly requiring that the corresponding first-order derivatives of the arm lengths be canceled out independently for both directions, one further picks out the modified second-generation TDI combinations from the remaining ones.
By definition, it will subsequently give rise to the modified second-generation TDI combinations.
According to geometric TDI, such an approach does not systematically remove all the second-order terms.
To be specific, the residual still consists of the contributions due to acceleration and the second-order ones in the spacecraft's velocity.
However, reminiscent of the difference between the first and modified first-generation TDI, we understand that it effectively cancels out a specific subset of second-order discrepancies.

In this work, we utilize the geometric TDI in conjunction with a ternary search algorithm to investigate the modified second-generation TDI combinations.
Using the method of exhaustion and eliminating redundant solutions, three, four, and forty second-generation TDI solutions are obtained respectively for twelve-, fourteen-, and sixteen-link combinations. 
Among the above solutions, we further separate, from the others, the combinations which are qualified to be the modified second-generation TDI.
For instance, nine of the sixteen-link combinations are recognized as of the latter type, already established in the literature~\cite{tdi-d88}, such as the Michelson, the Relay, the Beacon, and the Monitor combinations. 
There are no modified second-generation TDI solutions among the twelve- and fourteen-link combinations.
The properties of the modified second-generation TDI are analyzed, and we argue that some of these combinations, referred to as alternative combinations in this work, possess advantageous features and might be considered preferable alternatives.
In particular, the alternative combinations are essentially folded in time by inverse time-delay operations, and they are featured by reduced temporal footprint.
As pointed out in~\cite{tdi-geometric-2005}, the resultant contamination of the data set is likely to be minimized in the presence of instrumental gaps or glitches.
By analytically evaluating the response function and residual noise power spectral density (PSD), one may argue that the sixteen-link modified second-generation solutions are more favorable than the second-generation ones.
Although the noise suppression of both cases is of the same level, the magnitude of the gravitational wave signals is less pronounced in the case of the second generation TDI.
The latter can be attributed to the cancellation in the response function caused by the temporal symmetry of the arm length.
Besides, as a rather desirable feature, the response function of the alternative modified second-generation TDI possesses fewer zeros, consistent with the observations by Vallisneri~\cite{tdi-geometric-2005}.
Moreover, we derive the analytical expression for the leading-order residual mismatch between the two synthesized optical paths.
Such discrepancies are caused by the higher-order terms associated with the accelerations and products of spacecraft's velocities.
Their magnitudes are found to be at the order of pico-seconds.

The remainder of the paper is organized as follows.
In Sec.~\ref{section2}, the notations and conventions used in the paper are presented.
We review the basic notions of the TDI and geometric TDI, and the ternary search algorithm utilized in the study is discussed. 
Consequently, in Sec.~\ref{section3}, we present the results for the eight-link first-generation TDI, twelve-, fourteen-, and sixteen-link second-generation TDI combinations.
The modified second-generation TDI solutions are identified from the sixteen-link ones.
The analytical expressions of the optical path discrepancies are then derived in Sec.~\ref{section4}.
Sec.~\ref{section5} is dedicated to investigating the properties of the sensitivity function of the relevant TDI combinations.
The concluding remarks are given in Sec.~\ref{section6}.
The adopted index conventions are discussed in Appendix~\ref{apptidal}.
The corresponding Fourier coefficients that constitute the averaged response function are given in Appendix~\ref{apptida2}.

\section{geometric TDI and a ternary search algorithm}\label{section2}

\subsection{Definitions and conventions}\label{section2.1}

\begin{figure}[!t]
\includegraphics[width=0.40\textwidth]{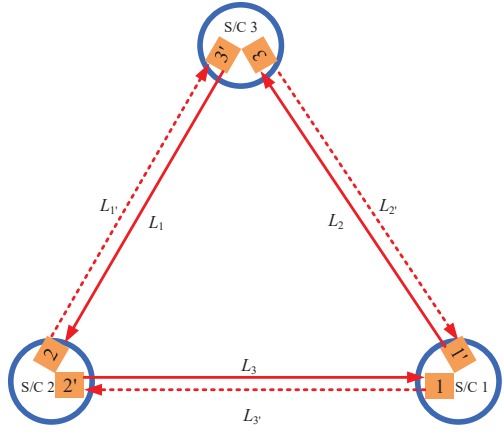}
\caption{\label{fig1} Notation defined in the space-based interferometric layout consists of GW detector, lasers and links.}
\end{figure}

For convenience, in this paper, we utilize subscript with a single index to indicate the physical quantities, such as the arm length and data stream~\cite{tdi-03}. 
The choice of the conventions is further discussed in Appendix~\ref{apptidal}.
The layout of the space-borne GW detector is illustrated in Fig.~\ref{fig1}. The constellation consists of three spacecraft (S/C), denoted by S/C $1$, S/C $2$, and S/C $3$ in the clockwise direction. 
An individual spacecraft carries two almost identical optical bench, represented by $i$ and $i'$ where $i=1,2,3$.
The two opposite armlengths of S/C $i$ are denoted by $L_i$ and $L_{i'}$, respectively, for counter-clockwise and clockwise propagation directions.
The information of the GW signals is primarily captured through the interference between the local and distant laser beams, known as the science data stream.
Based on the principle of three-segment laser interferometry~\cite{tdi-otto-2015}, the above measurement is further complemented by the test mass data stream and reference data stream.
In this study, we consider the laser frequency noise, test mass noise, and shot noise.

The data streams recorded at the optical benches ${i}$ and ${i'}$ are
\begin{align}\label{N1}
{s_i}(t) &= {\cal D}_{i - 1}p_{(i + 1)'}(t) - p_i(t) + {h_i}(t) + N_i^{opt}(t),\\\notag
{\varepsilon _i}(t) &= p_{i'}(t) - {p_i}(t) - 4\pi \nu_{i'}\vec n_{(i - 1)'} \vec \delta _{i}(t),\\\notag
{\tau _i}(t) &= p_{i'}(t) - p_i(t),
\end{align}
and
\begin{align}\label{N2}
s_{i'}(t) =& {\cal D}_{(i + 1)'} p_{i - 1}(t) - p_{i'}(t) + h_{i'}(t) + N_{i'}^{opt}(t),\\\notag
\varepsilon _{i'}(t) =& {p_i}(t) - p_{i'}(t) - 4\pi {\nu_i}\vec n_{i + 1} \vec \delta _{i'}(t),\\\notag
\tau _{i'}(t) =& {p_i}(t) - p_{i'}(t).
\end{align}
In the above equations, $s_i$ and $s_{i'}$ with ${i=1,2,3}$ and ${i'=1',2',3'}$ are the science data stream, which represent the interference between the laser beams from the local and distant spacecraft.
${\varepsilon _i}$ and $\varepsilon _{i'}$ are the test mass data stream, where the light beams are bounced off from test mass deliberately in order to capture its mechanical motion.
$\tau_i$ and $\tau _{i'}$ are the reference data stream, which represents the direct interference between two adjacent laser beams from the same spacecraft.
The possible GW signal, which demonstrates itself as a Doppler frequency shift, is embedded in the incident laser beam from the distant spacecraft. 
${h_i}$ and ${h_{i'}}$ represent such contributions due to the possible presence of a transverse-traceless GW.
${p_i}$ and $p_{i'}$ denote either the laser frequency noise or laser phase noise.
The latter two physical quantities, namely, the laser frequency and phase noise, are by and large equivalent,
but in the case of the laser frequency noise, there will be an additional Doppler shift, and the delay operator needs to be corrected (also see the discussions in Ref.~\cite{frequency-tdi-2021}).
In the remainder of the present paper, the calculations will be carried out by considering the laser frequency noise, which perturb around the center frequency ${\nu_{i }\approx 282~ \rm{THz}}$.
${N_i^{opt}}$ and ${N_{i'}^{opt}}$ are the shot noise,
${\vec \delta_i}$ and ${\vec \delta_{i'}}$ are associated with the mechanical vibrations of test mass with respect to the local inertial reference frame.
${{\cal D}_{i}}$, ${{\cal D}_{i'}}$ represent six time-delay operators.

The signal and noise embedded in the laser beam that propagates along a specific detector armlength $i$ are subjected to a time delay associated with the optical path connecting the two spacecraft.
Mathematically, the effect can be formulated by the action of a time-delay operator ${\cal D}_i$ on the relevant data streams ${x(t)}$.
To be specific, we have~\cite{tdi-otto-2015}
\begin{align}\label{N3}
{{\cal D}_j}x(t) =& x\left(t -\frac{1}{c}L_j(t)\right),\\\notag
{{\cal D}_i}{{\cal D}_j}x(t) = &{\cal D}_i x\left(t -\frac{1}{c}L_j(t)\right) = x\left(t-\frac{1}{c}L_i(t)-\frac{1}{c}L_j\left(t-\frac{1}{c}L_i(t)\right)\right),
\end{align}
where $c$ is the speed of light.
Successive applications of the time-delay operators can be simplified to
\begin{align}\label{N4}
{\cal D}_i {\cal D}_j x(t) \equiv {\cal D}_{i j}x(t).
\end{align}
To eliminate primed laser frequency noise and optical bench noise, one introduces intermediate variables~\cite{tdi-03,tdi-clock-Wang}.
The process gives rise to the following six observables
\begin{align}\label{eta}
{\eta _i}(t) = {h_i}(t) + {\cal D}_{i - 1} p_{i + 1}(t) - {p_i}(t) + 2\pi \nu _{(i + 1)'}{\vec n}_{i - 1}[ {\cal D}_{i - 1}\vec \delta_{(i + 1)'}(t) - \vec \delta {_i}(t)] + N_i^{opt}(t),\\\notag
\eta _{i'}(t) = h_{i'}(t) + {\cal D}_{(i + 1)'}p_{i - 1}(t) - {p_i}(t) + 2\pi \nu _{i - 1}\vec n_{i + 1}[ \vec \delta _{i'}(t) - {\cal D}_{(i + 1)'}\vec \delta _{i - 1}(t)] + N_{i'}^{opt}(t).
\end{align}
Moreover, the TDI algorithm constructs virtual equal armlength interference by linear combination of the above variables, namely~\cite{tdi-clock4},
\begin{align}\label{tdi}
{\rm TDI }= \sum\limits_{i = 1}^3 [P_i \eta _{i}(t) + P_{i'}\eta _{i'}(t)].
\end{align}
where ${P_{i}}$ and ${P_{i'}}$ are polynomials of the time-delay operators.
The laser frequency noise is expected to be eliminated from the above resultant expression.

\subsection{A ternary search for geometric TDI}\label{section2.2}

\begin{figure}[!t]
\includegraphics[width=0.20\textwidth]{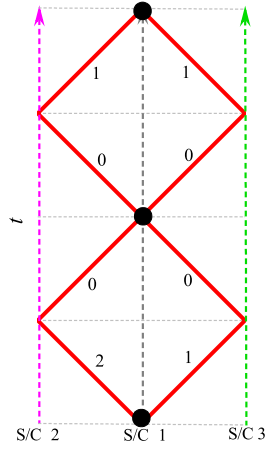}
\caption{\label{fig2} The space-time diagram of a geometic TDI solution for the the eight-link Michelson combination.}
\end{figure}

The geometric TDI is an algorithm that constitutes appropriate TDI combinations in the form of Eq.~\eqref{tdi} by the method of exhaustion~\cite{tdi-geometric-2005}.
For example, a typical geometric TDI solution is illustrated by the space-time diagram in Fig.~\ref{fig2}. 
The diagram is featured by discrete grids. 
The grids in the horizontal direction are spatial, and they correspond to the index numbers of the three spacecraft. While the vertical ones are temporal, the time propagation is upward.
A TDI combination can be seen as being composed of two optical paths, as illustrated by Fig.~\ref{fig2}.
Both paths start from a given S/C $i$ and eventually terminate at a second S/C $j$.
For instance, in Fig.~\ref{fig2}, the two optical paths start from S/C 1 and cease at S/C 1, which is the third intersection point from top to bottom.
Equivalently, one can visualize the TDI combination to construct a loop that eventually returns the starting spacecraft by propagating successively forward and backward in time.
To show the successive propagation of the laser between the spacecraft more transparently, when it is necessary, the grid indices can be periodically expanded in the horizontal direction (such as those shown in the diagrams Fig.~\ref{fig4} (c) and (d).)
An optical path may also propagate in the opposite direction of time, which gives rise to a time-advance operation.
Apparently, since the latter is the inverse action of the time delay, successive application of the time delay and time advance operations along the same armlength, known as the {\it null bigrams}, should not be considered due to its redundancy.
Subsequently, the algorithm exhausts all possible diagrams for a given number of total links and searches for valid solutions that satisfy the criterion discussed below.

Following the spirit of the geometric TDI, a valid TDI combination implies that the summations of optical paths along the two synthesized routes must be identical~\cite{tdi-geometric-2005}.
Equivalently, if one assign a negative sign if the the time propogation is backward, one has  
\begin{align}\label{geoTDIeq}
\sum_{\alpha} \vec{L}_\alpha(t_\alpha)  + \sum_{\beta} (-1)\cev{L}_\beta(t_\beta) = 0 ,
\end{align}
where the armlength $L_\alpha(t)$ is a function in time, 
$\alpha, \beta = 1, 2, 3, 1', 2', 3'$ that enumerates all the links that constitute the diagram, the superscripts ``$\rightarrow$'' and ``$\leftarrow$'' indicate the two forward and backward evolutions in time, and the summations are carried out for all the links on the optical paths.

The detector armlengths ${L}_i(t)$, as a function of time, can be rewritten using Taylor expansion about the instant of measurement as
\begin{align}\label{arm}
{L_i}(t) =& L_{i} + {{\dot L}_i}t + \frac{1}{2}{{\ddot L}_i}t^2 + \cdots \nonumber\\
{L_{i'}}(t) =& L_{i'} + {\dot L}_{i'}t + \frac{1}{2}{{\ddot L}_{i'}}{t^2} + \cdots
\end{align}
One substitutes the above equation into Eq.~\eqref{geoTDIeq}, groups up the terms proportional to the armlengths and their time derivatives, and divides both sides by $\frac{L}{c}$. 
The resulting delay-time residual along an closed optical path reads
\begin{align}\label{delat}
\frac{c\delta t}{L} = \sum\limits_{i = 1}^3\left(b_i\frac{L_i}{L} + b_{i'}\frac{L_{i'}}{L} + {d_i}\frac{\dot L_i}{c} + d_{i'}\frac{\dot L_{i'}}{c} +\frac{1}{2}{f_i}\frac{{\ddot L_i}L}{c^2}+ \cdots\right),
\end{align}
where ${L_i, L_{i'}}$ represent the armlengths, $\dot L_i,\dot L_{i'}$ are their first-order time derivative, $\ddot L_i$ correspond to the second order terms, all evaluated at a given instant.
One ignores the difference in the second order terms, so that $\ddot L_i\approx \ddot L_{i'} $. 
The dimensionless coefficients $b_i, b_{i'}$, $d_i, d_{i'}$, and $f_i$ are governed by the specific diagram.
The explicit forms of these coefficients are given, for instance, in Tab.~\ref{14linkTab} for the second-generation fourteen-link diagrams and Tab.~\ref{16linkTab} for the modified second-generation sixteen-link ones in the following section. 
In deriving Eq.~\eqref{delat}, one also makes use of the approximation $L = c T $ where $L\sim L_i$.
Therefore, $\delta t$ essentially contains the contributions of second-order terms, which are not considered in the first and second TDI combinations.
The remaining higher-order contributions from the Taylor expansion and those from the above approximations are insignificant and thus attributed to ``$\cdots$''.

The criterion for the first generation TDI is derived by only taking into account the leading terms in the expansion Eq.~\eqref{delat}.
In particular, the first generation TDI is obtained by requiring 
\begin{align}\label{geoTDIFirst}
b_i + b_{i'}=0 .
\end{align}
On the other hand, the modified first generation TDI demands that
\begin{align}\label{geoTDImodFirst}
b_i =  b_{i'}=0 .
\end{align}
Apparently, the first generation TDI gives a rigorous solution if the detector constellation is static, namely, $L_i(t) = L_{i'}(t) \equiv L_i$. 
The modified first-generation TDI serves as a solution if the constellation is rigid and rotates at a constant speed.
This is because, for the latter case, one effectively has constant armlengths $L_i(t) = L_i \ne L_{i'} = L_{i'}(t)$ due to the symmetry in time translation.

For the second generation TDI, one further includes the first-order derivative terms.
The criterion for the second generation reads
\begin{align}\label{geoTDISecond}
b_i = b_{i'}=0 ,\nonumber\\
d_i + d_{i'}=0 .
\end{align}
However, generally one has $\dot L_i \ne \dot L_{i'}$, and some discrepancy from $\dot L_i=\dot L_{i'}$ is expected (also see the explicit form given by Eq.~\eqref{LdotNewDef} below).
In other words, the most dominant contribution on the r.h.s. of Eq.~\eqref{delat} gives
\begin{align}\label{delaytt}
\frac{c\delta t}{L}  \approx \sum\limits_{i = 1}^3 {d_i}\frac{\dot L_i - \dot L_{i'}}{c}+\sum\limits_{i = 1}^3 \frac{1}{2}{f_i}\frac{{\ddot L_i}L}{c^2}.
\end{align}

Therefore, following the line of thought of the modified first-generation TDI, one may distinguish the rates of change of arm lengths in the two opposite directions.
This gives rise to the modified second generation TDI, which satisfies
\begin{align}\label{geoTDImodSecond}
b_i = b_{i'}=0 ,\nonumber\\
d_i = d_{i'}=0 .
\end{align}
In this case, one is free of the problem caused by the speed mismatch $\dot L_i\ne \dot L_{i'}$. Now, Eq.~\eqref{delat} gives
\begin{align}\label{delaytt2}
\frac{c\delta t}{L} \approx \sum\limits_{i = 1}^3 \frac{1}{2}{f_i}\frac{{\ddot L_i}L}{c^2}.
\end{align}
which is essentially of second order in derivatives, and is expected to be of minor magnitude. 
Eqs.~\eqref{delaytt} and~\eqref{delaytt2} will be utilized to estimate the optical path descrepancies and explicitly evaluated in Sec.~\ref{section4}.

\begin{figure}[!t]
\includegraphics[width=0.40\textwidth]{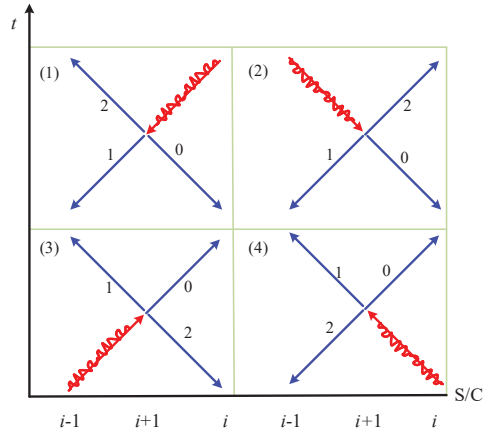}
\caption{\label{fig3} The four possible space-time diagrams where one considers the preceding spacecraft from which the laser beam was deflected.
For all the cases, one of the four propagation schemes, indicated by the wavy line, is identified to be a null bigram and, therefore, can be excluded from the enumeration process.}
\end{figure}

We now turn to discuss the enumeration process of the geometric TDI.
In Ref.~\cite{tdi-geometric-2005}, the search was performed by enumerating a total of ${4^n}$ possible diagrams with $n$ links.
However, the computationally expensive task can be somewhat relieved by the following considerations.
It is noted that both the laser propagation and the time evolution directions can be denoted by a binary digit, and subsequently, $2\times 2=4$, there are four possibilities.
To be specific, one adopts the convention that the clockwise propagations $\rm{S/C~3} \to S/C~1 \to S/C~2 \to S/C~3$ are denoted by 1, and the counterclockwise propagations $\rm{S/C~3}\to S/C~2 \to S/C~1 \to S/C~3$ by 0.
Also, the laser beam that propagates forward in time is denoted by ``1'' and backward in time by ``0''.
As explicitly shown in Fig.~\ref{fig3}, as long as one considers the preceding spacecraft where the laser beam was deflected, among the above four possibilities, one of them (indicated by the wavy line) is always a null bigram and can readily be excluded.
As a result, apart from the first link, there are only three relevant possibilities for laser propagation.
Therefore, a ternary search algorithm is adopted to look for the feasible TDI combinations, with a total of $3^n$ possible diagrams with $n$ links, which significantly reduces the computational load.
The three possible optical paths are denoted by ``0'', ``1'', and ``2''.
In particular, the two choices with the same time evolution direction are denoted by ``0'' and ``1'' according to the cyclic order for the spatial armlength. The remaining one is denoted by ``2''.
Also, without loss of generality, one assumes that the index of the start spacecraft $i=1$ and the first link is ``1''. 
For instance, the eight-link Michelson combination shown in Fig.~\ref{fig2} is thus written as $\left\{10012001\right\}$.

Consider the case of $n$-link combinations as an example. 
The enumeration process start from the diagram $\left\{ 1,\underbrace {0,...0}_{n-1} \right\}$, which corresponds to a ternary number with the value $n_{\mathrm{min}} = 3^{n - 1}$.
It ceases at the diagram $\left\{ 1,\underbrace {2,...2}_{n-1} \right\}$ with the the maximum value $n_{\mathrm{max}} =  2 \times 3^{n - 1} - 1$.
The difference $n_{\mathrm{max}} - n_{\mathrm{min}} + 1 = 3^{n - 1}$ gives the number of possible diagrams.
By exhausting all these diagrams using the criterion given by Eqs.~\eqref{geoTDIFirst}-\eqref{geoTDImodSecond}, it is necessary to check whether the obtained solutions are redundant.

For the completeness of the discussion, we briefly mention why the geometric TDI solution satisfying Eq.~\eqref{geoTDIeq} furnishes a valid TDI solution of Eq.~\eqref{tdi}.
According to Ref.~\cite{tdi-geometric-2005}, one assigns the corresponding observable Eq.~\eqref{eta} to each node of a geometric TDI solution and then sums up individual contributions from all the grids with appropriate time delays.
A valid solution thus implies that all the terms regarding the three independent laser frequency noise cancel out entirely.
It is apparent that Eq.~\eqref{geoTDIeq} dictates that the laser frequency noise $p_i$ of the starting S/C $i$ cancel out.
This is because the two optical paths through the two virtual routes are identical when they terminate at S/C $j$, and subsequently, the time delays along the two routes are the same.
However, due to the specific form of Eq.~\eqref{eta}, the first links of both routes also inevitably brings in an additional noise term originating from the receiving spacecraft of the link multiplied by a minus sign.
It is noted that this additional term is eliminated by the corresponding observable $\eta_i$ assigned to this grid since it is readily verified that both terms experience the same amount of time delay and thus are identical up to a minus sign.
One may verify that such cancellations can be carried out iteratively until the grid where the two routes meet.
Neither of the two laser frequency noise terms $p_j$ from the two corresponding links is subjected to any time delay, and therefore all the noise terms are annihilated.

\section{The resulting geometric TDI solutions}\label{section3}

This section presents the resulting geometric TDI combinations generation, obtained by exhausting the eight-, twelve-, fourteen-, and sixteen-link diagrams.
Our focus is on the modified versions of the TDI solutions, for which the differences between $L_i$ and $L_{i'}$ and their time-derivatives are taken into consideration explicitly.
As it turns out, nine sixteen-link solutions belong to the modified second-generation combinations, which include the ones previously derived in the literature~\cite{tdi-d88}, such as the Michelson, the Relay, the Beacon, and the Monitor combinations. 
We refer to the remaining ones as {\it alternative} solutions and study their properties.
 
\subsection{Eight-link modified first-generation TDI combinations}

\begin{figure}[!t]
\includegraphics[width=0.60\textwidth]{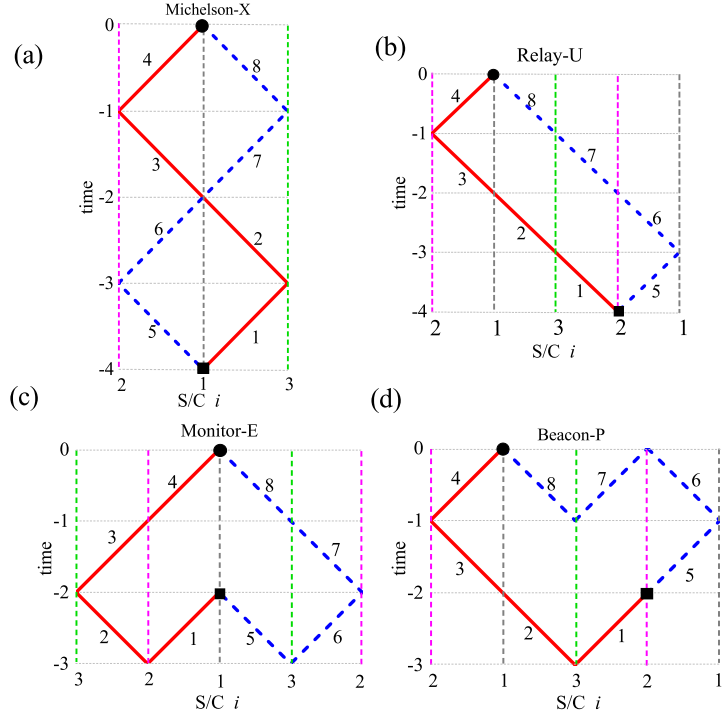}
\caption{\label{fig4}  
The space-time diagrams for the eight-link TDI combinations: Michelson-$X$, Relay-$U$, Monitor-$E$, and Beacon-$P$.
The red-solid and blue-dashed line segments correspond to the two synthesized routes, either of which might contain both forward and backward propagations in time.
The black squares and black dots indicate the initial and terminal grids of the two routes, and the reference time $t=0$ is attributed to the terminal S/C 1.}
\end{figure}

For the first-generation TDI combinations, one assumes that the six armlengths satisfy $L_i = L_{i'}={\rm{const}}.$, but $L_1\ne L_2\ne L_3$. 
The modified first-generation TDI combinations further assumes that the six armlengths are distinct, namely, $L_i(t) = L_i \ne L_{i'} = L_{i'}(t)$. 
There are only four modified first-generation TDI combinations, and their space-time diagrams of the geometric TDI solutions are illustrated in Fig.~\ref{fig4}. 
The red-solid and blue-dashed line segments correspond to the two synthesized routes, either of which might contain both forward and backward propagations in time.
The initial and terminal grids of the two routes are indicated by the black squares and black dots, respectively.
Also, without loss of generality, the reference time $t=0$ is assigned to the terminal S/C 1.
In this study, the above convention is adopted by the remaining plots.

To be specific, following the convention developed in~\cite{tdi-geometric-2020}, the time series of Figs.~\ref{fig4}~(a), (b), (c) and (d) are illustrated in Tab.~\ref{8linkTrajectory}.
\begin{table}
\centering
\caption{Trajectory of the eight-link modified first-generation TDI combination laser link, where, 1, 2, and 3 represents the indices of the spacecraft,
$1 \to 2$  indicates that the laser emitted from S/C 1 propogates to S/C 2.
Similarly, $1 \leftarrow 2$ means that the laser emitted from S/C 2 propagates to S/C 1 in the positive direction of time, or the laser propagates from S/C 1 to S/C 2 in the opposite direction of time.}
\newcommand{\tabincell}[2]{\begin{tabular}{@{}#1@{}}#2\end{tabular}}
\renewcommand\arraystretch{2}
\begin{tabular}{|c|c|}
 \hline
 \hline
\makecell[c] {TDI\\ combination} & Laser link trajectory \\
 \hline
$X(t)$&$ \!1 \! \leftarrow  \!2  \! \leftarrow  \!1 \! \leftarrow  \!3 \! \leftarrow  \!1 \! \to \! 2 \! \to  \!1 \! \to  \!3 \! \to  \!1 \!$\\
\hline
$U(t)$ & $ \!1 \! \leftarrow  \!2 \! \leftarrow  \!1 \! \leftarrow  \!3 \! \leftarrow  \!2 \! \to  \!1 \! \to  \!2 \! \to  \!3 \!\to  \!1 \!$\\
\hline
$E(t)$&$ \!1 \! \leftarrow 2 \leftarrow \! 3 \! \leftarrow  \!2 \!\to  \!1 \! \leftarrow \! 3 \!\to  \!2 \! \to  \!3 \! \to  \!1 \!$\\
\hline
$P(t)$&$ \!1  \!\leftarrow \! 2  \!\leftarrow  \!1 \! \leftarrow  \!3  \!\to  \!2 \!\to  \!1 \! \to  \!2 \!\leftarrow  \!3 \! \to  \!1 \!$\\
\hline
\hline
\end{tabular}
\label{8linkTrajectory}
\end{table}

Accordingly, the resulting TDI combinations are found to be consistent with the existing literature~\cite{tdi-03}.    
For Fig.~\ref{fig4}(a),
\begin{align}\label{X1}
X(t) ={\cal D}_{33'2'}\eta _3+  {\cal D}_{33'} \eta _{1'}+ {\cal D}_3 \eta _{2'} +\eta _1 - \left({\cal D}_{2'23} \eta _{2'}+ {\cal D}_{2'2} \eta _1+ {\cal D}_{2'} \eta _3 + \eta _{1'}   \right).
\end{align}
It corresponds to the Michelson combination.
The first four terms of Eq.~\eqref{X1} correspond to the virtual optical paths via the links $1-2-3-4$, and the last four terms correspond to the one in terms of the links $5-6-7-8$. 

Fig.~4(b) gives the Relay TDI combination, which reads
\begin{align}\label{U1}
U(t) ={\cal D}_{33'2'} \eta _{3'} +{\cal D}_{33'} \eta _{1'}+ {\cal D}_3 \eta _{2'}+\eta _1- \left({\cal D}_{2'1'3'} \eta _1 + {\cal D}_{2'1'} \eta _{2'}+ {\cal D}_{2'}\eta _{3'}+  \eta _{1'}\right).
\end{align}

The first four terms of Eq.~\eqref{U1} correspond to the virtual optical paths via the links $1-2-3-4$, and the last four terms correspond to the one in terms of the links $5-6-7-8$. 
Based on the cycle rules between the three spacecraft, the Relay TDI combination contains $U(t),V(t),W(t)$,
where $U(t)$ was denoted as $W(t)$ in some literatures~\cite{tdi-03,tdi-otto-2015}.

Fig.~4(c) gives the Monitor TDI combination, which reads
\begin{align}\label{E1}
E(t) =  - {\cal D}_{311'\bar 3} \eta _1+ {\cal D}_{31} \eta _{3'}+ {\cal D}_3 \eta _2+\eta _1- \left(- {\cal D}_{2'1'1\bar 2'}\eta _{1'}+ {\cal D}_{2'1'}\eta _2+ {\cal D}_{2'}\eta _{3'}+ \eta _{1'}\right),
\end{align}
where ${\cal D}_{\bar 3}$ is an advance operator which is the inverse of the corresponding delay operator, 
satisfying ${\cal D}_{\bar 3}{\cal D}_3 = {\cal D}_3 {\cal D}_{\bar 3} = 1$.
The first four terms of Eq.~\eqref{E1} correspond to the virtual optical paths via the links $1-2-3-4$, and the last four terms correspond to the one in terms of the links $5-6-7-8$. 
Because the modified first-generation TDI combination ignores the commutator of time-delay operators, the Eq.~\eqref{E1} can be further simplified to
\begin{align}\label{EEE1}
E(t)= \eta _1 + {\cal D}_3 \eta _2 + {\cal D}_{31} \eta _{3'} + {\cal D}_{1'1} \eta _{1'}- \left( \eta _{1'} + {\cal D}_{2'} \eta _{3'} + {\cal D}_{2'1'} \eta _2 + {\cal D}_{11'} \eta _1 \right).
\end{align}

Fig.~4(d) gives the Beacon TDI combination. The combination $P(t)$ is different from the other three given above.
It consists of two independent virtual optical loops.
For the space-time diagram, we choose $t=0$ and S/C 1 as the final grid for both loops, and one finds:
\begin{align}\label{P1}
P(t) = ( - {\cal D}_{33'2'\bar1} \eta _{2} + {\cal D}_{33'}\eta _{1'}+ {\cal D}_3 \eta _{2'}+ \eta _1 ) - ({\cal D}_{2' \bar 1 3'} \eta _1+ {\cal D}_{2' \bar 1} \eta _{2'} -{\cal D}_{2'\bar 1} \eta _2 + \eta _{1'} ).
\end{align}
The expression in literatures~\cite{tdi-03,tdi-otto-2015} is
\begin{align}\label{PP1}
P(t) = {\cal D}_1 \eta _1 + {\cal D}_{13} \eta _{2'} + {\cal D}_{133'} \eta _{1'} + {\cal D}_{2'}\eta _2- \left( {\cal D}_1\eta _{1'} + {\cal D}_{2'}\eta _{2'} + {\cal D}_{2'3'} \eta _1 + {\cal D}_{2'33'} \eta _2 \right).
\end{align}
The time delay operator ${\cal D}_1$ acts on the Eq.~\eqref{P1} given by the geometric space-time diagram, which is the same as the mathematical Eq.~\eqref{PP1}.

It is noted that the four TDI combinations given above do not form a complete generating set for an arbitrary modified first-generation TDI combination.
By adding two cyclic combinations to these four combinations, one furnishes a complete basis~\cite{tdi-geome-2002,tdi-laser-04}.

\subsection{Twelve-link second-generation TDI combinations}

\begin{figure}[!t]
\includegraphics[width=0.50\textwidth]{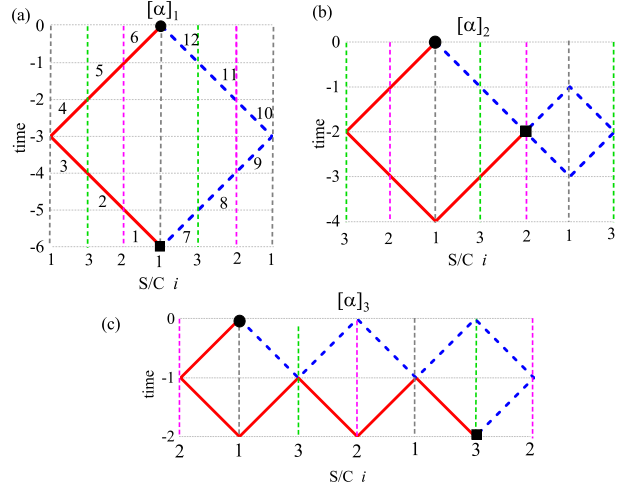}
\caption{\label{fig5} The space-time diagrams for the twelve-link geometric TDI combinations.}
\end{figure}

Using the criterion given by Eq.~\eqref{geoTDISecond} to study the twelve-link diagrams, one finds three geometric TDI combinations, as shown in  Fig.~\ref{fig5}.
The corresponding time series are illustrated in Tab.~\ref{12linkTrajectory}.
\begin{table}
\centering
\caption{Trajectory of the twelve-link second-generation TDI combination laser link.}
\newcommand{\tabincell}[2]{\begin{tabular}{@{}#1@{}}#2\end{tabular}}
\renewcommand\arraystretch{2}
\begin{tabular}{|c|c|}
\hline
\hline
\makecell[c] {TDI\\ combination} & Laser link trajectory \\
 \hline
$\left[ \alpha  \right]_1$&$1 \!\!\leftarrow \!\! 2\!\! \leftarrow \!\! 3 \leftarrow \!\!1\!\! \leftarrow \!\! 3 \!\! \leftarrow \!\!2 \!\!\leftarrow \!\! 1 \!\! \to \! 3 \!\! \to \!\! 2 \!\! \to \!\! 1 \!\! \to \!\!2 \!\! \to \!\! 3\! \!\to \!\!1$\\
\hline
$\left[ \alpha  \right]_2$ & $1 \!\!\leftarrow \!\!2 \!\!\leftarrow \!\!3 \leftarrow \!\!2 \!\!\leftarrow\! \!1 \!\! \to \!\!3 \to \!\!  2 \!\!\to\! 1\! \!\leftarrow \!\! 3 \!\! \leftarrow \!\!1 \! \!\to \!\! 2 \!\! \to\! \!3 \! \!\to \! \!1$\\
\hline
$\left[ \alpha  \right]_3$&$1 \!\!\leftarrow \!\! 2\!\! \leftarrow \!\! 1 \to \!\!3 \!\!\leftarrow \!\! 2 \!\! \to \!\!1\!\! \leftarrow \!\! 3 \!\!\to \!\! 2 \! \!\to \!\!3 \!\! \leftarrow \!\!1 \! \!\to \!\! 2 \!\! \leftarrow \!\! 3\! \! \to\! \! 1$\\
\hline
\hline
\end{tabular}
\label{12linkTrajectory}
\end{table}

It is noted that $\left[ \alpha  \right]_1$ is a known form of the second-generation TDI combination, which reads
\begin{align}\label{alpha1}
\left[ \alpha  \right]_1 = {\cal D}_{3122'1'}\eta _{2'}\!\! +\!\!  {\cal D}_{3122'}\eta _{3'}\!\! +\!\!  {\cal D}_{312} \eta _{1'} \!\! +\!\!  {\cal D}_{31} \eta _3 \!\! +\!\!  {\cal D}_3 \eta _2 \!\! +\!\! \eta _1 -({\cal D}_{2'1'3'31} \eta _3  \!\!+\!\! {\cal D}_{2'1'3'3} \eta _2\!\!+\!\! {\cal D}_{2'1'3'} \eta _1 \!\!+\!\! {\cal D}_{2'1'} \eta _{2'} \!\!+\!\! {\cal D}_{2'} \eta _{3'} \!\!+\!\!\eta _{1'} ).
\end{align}

The first six terms of the Eq.~\eqref{alpha1} correspond to the virtual optical paths via the links $1-2-3-4-5-6$, and the last six terms correspond to the one in terms of the links $7-8-9-10-11-12$. 
Comparing with the optical path with $\left[ \alpha  \right]_1$, $\left[ \alpha  \right]_2$, and $\left[ \alpha  \right]_3$,
it can be found that the $\left[ \alpha  \right]_2$, $\left[ \alpha  \right]_3$ combinations of alternative forms spans shorter time in the time domain.
Therefore, the alternative TDI combinations are not easily affected by instrumental gaps or glitches.
On the other hand, the alternative TDI combination $[\alpha]_2$ has better sensitivity in the high-frequency region.
This is because the averaged response function and noise PSD have fewer zeros, as shown by Figs.~\ref{fig19} and~\ref{fig20} of Sec.~\ref{section5}.

We give in the second and third columns of Tab.~\ref{12linkTab}, the coefficients $d_i, d_{i'}$ and $f_i$ defined earlier by Eq.~\eqref{delat} of the twelve-link 2nd-generation geometric TDI combinations. 
These quantities are then utilized to eventually evaluate the residual $\frac{c\delta t}{L}$ and shown by the fourth column of the table (see the derivations given in Sec.~\ref{section4}).

\begin{table}
\caption{Delay-time residual of the twelve-link geometric TDI combinations. 
The fifth and sixth columns were evaluated using the parameters of the LISA detector.}
\centering
\newcommand{\tabincell}[2]{\begin{tabular}{@{}#1@{}}#2\end{tabular}}
\renewcommand\arraystretch{2}
\begin{tabular}{|c|c|c|c|c|c|}
\hline
\hline
\makecell[c] {TDI\\ combination} & $\{d_i,d_{i'}\}$ &$ \{f_i\}$ &time residual $\frac{c\delta t}{L}$&\makecell[c] {coefficient\\ before $\sin3\Omega t$ (s)}&\makecell[c] {coefficient\\ before $\cos\Omega t$ (s)}\\
 \hline
 $[\alpha]_1$ &$\{3,3,3,-3,-3,-3\}$ & $\{0,-6,6\}$ &$3\sum\limits_{i = 1}^3 \frac{\dot{L}_i - \dot{L}_{i'}}{c}+6\frac{L}{c^2}(\ddot L_3-\ddot L_2)$&$2.7\times10^{-13}$&$2.7\times10^{-13}$\\
 \hline
 $[\alpha]_2$ &$\{1,1,1,-1,-1,-1\}$& $\{0,-2,2\}$ & $\sum\limits_{i = 1}^3 \frac{\dot{L}_i- \dot{L}_{i'}}{c}+2\frac{L}{c^2}(\ddot L_3-\ddot L_2)$&$9\times10^{-14}$&$9\times10^{-14}$\\
 \hline
 $[\alpha]_3$ &$\{1,1,1,-1,-1,-1\}$& $\{0,0,0\}$ & $\sum\limits_{i = 1}^3 \frac{\dot{L}_i - \dot{L}_{i'}}{c}$&$9\times10^{-14}$&$0$\\
 \hline
\hline
\end{tabular}
\label{12linkTab}
\end{table}

\subsection{Fourteen-link second-generation TDI combinations}

\begin{figure}[!t]
\includegraphics[width=0.60\textwidth]{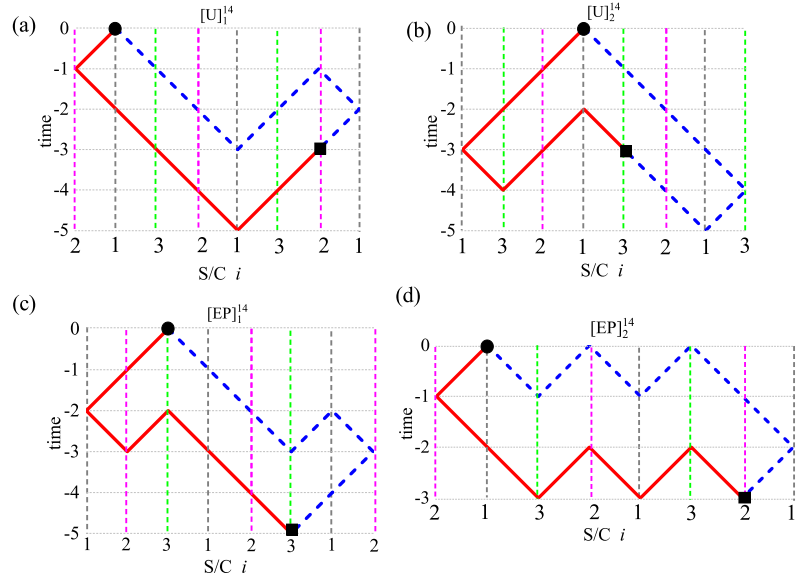}
\caption{\label{fig6} The space-time diagrams for the  fourteen-link geometric TDI combinations.}
\end{figure}

Similarly, by employing the criterion given by Eq.~\eqref{geoTDISecond} to study the fourteen-link diagrams, one finds four geometric TDI combinations, as shown in Fig.~\ref{fig6}.
The corresponding time series are given in Tab.~\ref{14linkTrajectory}.

\begin{table}
\centering
\caption{Trajectory of the fourteen-link second-generation TDI combination laser link.}
\newcommand{\tabincell}[2]{\begin{tabular}{@{}#1@{}}#2\end{tabular}}
\renewcommand\arraystretch{2}
\begin{tabular}{|c|c|}
\hline
\hline
\makecell[c] {TDI\\ combination} & Laser link trajectory \\
 \hline
$[U]_1^{14}$&$1 \leftarrow 2 \leftarrow 1 \leftarrow 3 \leftarrow 2 \leftarrow 1 \to 3 \to 2 \to 1 \to 2 \leftarrow 3 \leftarrow 1 \to 2 \to 3 \to 1$\\
\hline
$[U]_2^{14}$ & $1 \leftarrow 2 \leftarrow 3 \leftarrow 1 \leftarrow 3 \to 2 \to 1 \leftarrow 3 \leftarrow 2 \leftarrow 1 \to 3 \to 1\to 2 \to 3 \to 1$\\
\hline
$[EP]_1^{14}$&$  1 \leftarrow 2 \leftarrow 3 \leftarrow 2 \to 1 \leftarrow 3 \leftarrow 2 \leftarrow 1 \to 3 \to 2 \to 3 \leftarrow 1 \to 2 \to 3 \to 1$\\
\hline
$[EP]_2^{14}$&$ 1 \leftarrow 2 \leftarrow 1 \leftarrow 3 \to 2 \leftarrow 1 \to 3 \leftarrow 2 \to 1 \to 2 \to 3 \leftarrow 1 \to 2 \leftarrow 3 \to 1$\\
\hline
\hline
\end{tabular}
\label{14linkTrajectory}
\end{table}

\begin{table}
\caption{Delay-time residual of the fourteen-link geometric TDI combinations. 
The fifth and sixth columns were evaluated using the parameters of the LISA detector.}
\centering
\newcommand{\tabincell}[2]{\begin{tabular}{@{}#1@{}}#2\end{tabular}}
\renewcommand\arraystretch{2}
\begin{tabular}{|c|c|c|c|c|c|}
\hline
\hline
\makecell[c] {TDI\\ combination} & $\{d_i,d_{i'}\}$ &$ \{f_i\}$ &time residual $\frac{c\delta t}{L}$&\makecell[c] {coefficient\\ before $\sin3\Omega t$ (s)}&\makecell[c] {coefficient\\ before $\cos\Omega t$ (s)}\\
 \hline
 $[U]_1^{14}$ &$\{2,2,2,-2,-2,-2\}$& $\{0,-4,4\}$&$2\sum\limits_{i = 1}^3 \frac{\dot{L}_i - \dot{L}_{i'}}{c}+4\frac{L}{c^2}(\ddot L_3-\ddot L_2)$&$1.8\times10^{-13}$&$1.8\times10^{-13}$\\
 \hline
 $[U]_2^{14}$ &$\{2,2,2,-2,-2,-2\}$& $\{0,-4,4\}$& $2\sum\limits_{i = 1}^3 \frac{\dot{L}_i - \dot{L}_{i'}}{c}+4\frac{L}{c^2}(\ddot L_3-\ddot L_2)$&$1.8\times10^{-13}$&$1.8\times10^{-13}$\\
 \hline
 $[EP]_1^{14}$ &$\{2,2,2,-2,-2,-2\}$& $\{0,-4,4\}$& $2\sum\limits_{i = 1}^3 \frac{\dot{L}_i - \dot{L}_{i'}}{c}+4\frac{L}{c^2}(\ddot L_3-\ddot L_2)$&$1.8\times10^{-13}$&$1.8\times10^{-13}$\\
 \hline
 $[EP]_2^{14}$ &$\{2,2,2,-2,-2,-2\}$& $\{0,0,0\}$& $2\sum\limits_{i = 1}^3 \frac{\dot{L}_i - \dot{L}_{i'}}{c}$&$1.8\times10^{-13}$&$0$\\
 \hline
\hline
\end{tabular}
\label{14linkTab}
\end{table}

One observes that the resulting geometric TDI solutions can be expressed in the modified first-generation TDI combination.
In particular, $[U]_1^{14},[U]_2^{14}$ are the summation of two modified first-generation Relay combinations.
$[EP]_1^{14},[EP]_2^{14}$ are the summation of a modified first-generation Monitor combination and a modified first-generation Beacon combination. 
In order to distinguish it from the sixteen-link TDI combinations introduced later, these combinations are denoted by the superscript ``14''.

We enumerate in the second and third columns of Tab.~\ref{14linkTab}, the coefficients $d_i, d_{i'}$ and $f_i$ defined earlier by Eq.~\eqref{delat} of the fourteen-link 2nd-generation geometric TDI combinations.
These quantities are then utilized to eventually evaluate the residual $\frac{c\delta t}{L}$ and shown by the fourth column of the table (see the derivations given in Sec.~\ref{section4}).

\subsection{Sixteen-link modified second-generation TDI combinations}

Using the criterion given by Eq.~\eqref{geoTDISecond} to study the sixteen-link diagrams, one finds fourty geometric TDI combinations as shown in Tab~\ref{16linkTrajectory1},~\ref{16linkTraje2},~\ref{16linkTra3}.
Using the criterion given by Eq.~\eqref{geoTDImodSecond} to study the sixteen-link diagrams, we found nine solutions as the modified second-generation TDI combinations. 
They are given in Tab~\ref{16linkTrajectory1}.

\begin{table}
\centering
\caption{Trajectory of the sixteen-link modified second-generation TDI combination laser link.}
\newcommand{\tabincell}[2]{\begin{tabular}{@{}#1@{}}#2\end{tabular}}
\renewcommand\arraystretch{2}
\begin{tabular}{|c|c|}
\hline
\hline
TDI combination & Laser link trajectory \\
 \hline
$[X]_1^{16}$&$1 \leftarrow 2 \leftarrow 1 \leftarrow 3 \leftarrow 1 \leftarrow 3 \leftarrow 1 \leftarrow 2 \leftarrow 1 \to 3 \to 1 \to 2 \to 1 \to 2 \to 1 \to 3 \to 1$\\
\hline
$[X]_2^{16}$ & $1 \leftarrow 2 \leftarrow 1 \leftarrow 3\leftarrow 1 \leftarrow 2 \leftarrow 1 \to 3 \to 1 \to 2 \to 1 \leftarrow 3 \leftarrow 1 \to 2 \to 1 \to 3 \to 1$\\
\hline
$[U]_1^{16}$&$1 \leftarrow 2 \leftarrow 1 \leftarrow 3 \leftarrow 2 \to 1 \leftarrow 3 \leftarrow 2 \leftarrow 1 \leftarrow 2 \to 3 \to 1 \to 2 \to 1 \to 2 \to 3 \to 1$\\
\hline
$[U]_2^{16}$&$1 \leftarrow 2 \leftarrow1 \leftarrow 3 \leftarrow 2 \leftarrow 1 \leftarrow 2 \to 3 \to 1 \to 2 \to 1 \leftarrow 3 \leftarrow 2 \to 1 \to 2 \to 3 \to 1$\\
\hline
$[U]_3^{16}$&$1 \leftarrow 2 \leftarrow 1 \leftarrow 3 \leftarrow 2 \to 1 \to 2 \to 1 \leftarrow 3 \leftarrow 2 \leftarrow 1 \leftarrow 2 \to 3 \to 1 \to 2 \to 3 \to 1$\\
\hline
$[E]_1^{16}$&$1 \leftarrow 2 \leftarrow 3 \leftarrow 2 \to 1 \leftarrow 3 \leftarrow 2 \leftarrow 3 \to 1 \leftarrow 2 \to 3 \to 2 \to 1 \leftarrow 3 \to 2 \to 3 \to 1$\\
\hline
$[E]_2^{16}$&$1 \leftarrow 2 \leftarrow 3 \leftarrow 2 \leftarrow 3 \to 1 \leftarrow 2 \to 3 \to 2 \to 1 \leftarrow 3 \leftarrow 2 \to 1 \leftarrow 3 \to 2 \to 3 \to 1$\\
\hline
$[P]_1^{16}$&$1 \leftarrow 2 \leftarrow 1 \leftarrow 3\to 2 \leftarrow 1 \leftarrow 2 \leftarrow 3 \to 1 \to 2 \to 1 \leftarrow 3 \to 2 \to 1 \to 2 \leftarrow 3 \to 1$\\
\hline 
$[P]_2^{16}$&$1\leftarrow 2 \leftarrow 1 \leftarrow 3 \to 2 \to 1 \leftarrow 3 \to 2 \leftarrow 1 \leftarrow 2 \leftarrow 3 \to 1 \to 2 \to 1 \to 2 \leftarrow 3 \to 1$\\
\hline
\hline
\end{tabular}
\label{16linkTrajectory1}
\end{table}

\begin{figure}[!t]
\includegraphics[width=0.30\textwidth]{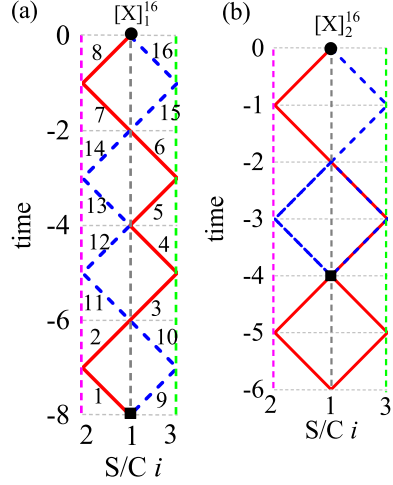}
\caption{\label{fig7} The space-time diagrams for the sixteen-link Michelson-type geometric TDI combinations.}
\end{figure}

When compared with the existing standard TDI combinations in literature~\cite{tdi-03}, two of the solutions, $[X]_1^{16},[X]_2^{16}$ are of Michelson-type.
The corresponding space-time diagram is illustrated in Fig.~\ref{fig7}.
It is observed that $[X]_1^{16}$ is identical to the traditional form Michelson combination in the literature~\cite{tdi-03,tdi-otto-2015}, which reads
\begin{align}\label{X2}
[X]_1^{16} =& {\cal D}_{33'2'22'23} \eta _{2'}  + {\cal D}_{33'2'22'2} \eta _1 + {\cal D}_{33'2'22'} \eta _3  + {\cal D}_{33'2'2} \eta _{1'}+ {\cal D}_{33'2'} \eta _3+ {\cal D}_{33'} \eta _{1'}+ {\cal D}_3 \eta _2+ \eta _1\\\notag
 -& \left( \eta _{1'} + {\cal D}_{2'} \eta _3 + {\cal D}_{2'2} \eta _1 + {\cal D}_{2'23} \eta _{2'} + {\cal D}_{2'233'} \eta _1 + {\cal D}_{2'233'3} \eta _{2'} + {\cal D}_{2'233'33'} \eta _{1'}+ {\cal D}_{2'233'33'2'} \eta _3 \right).
\end{align}

The first row of Eq.~\eqref{U1} corresponds to the virtual optical paths via the links $1-2-3-4-5-6-7-8$, and the second row corresponds to the one in terms of the links $9-10-11-12-13-14-15-16$. 
Similarly, the space-time diagram of the combination $[X]_2^{16}$ is also shown in Fig.~\ref{fig7}(b). 

\begin{figure}[!t]
\includegraphics[width=0.70\textwidth]{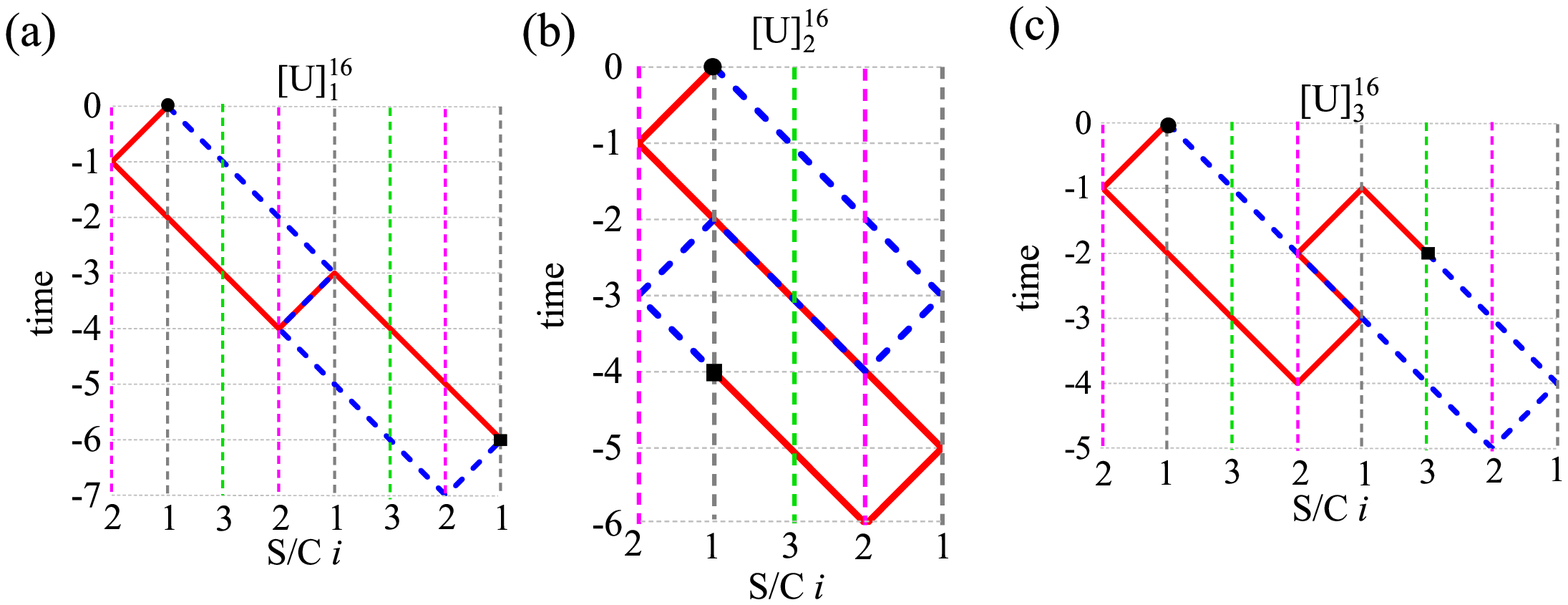}
\caption{\label{fig8}  The space-time diagrams for the sixteen-link Relay-type geometric TDI combinations.}
\end{figure}

To determine whether an obtained geometric TDI solution is equivalent to an existing one in the literature, one may repeat the process to construct the solution based on their modified first-generation counterpart.
Accordingly, it can be shown that the combinations $[U]_1^{16}, [U]_2^{16}, [U]_3^{16}$ are Relay-type TDI, and their space-time diagrams are illustrated in Fig.~\ref{fig8}.
To be specific, by using Fig.~\ref{fig8}(a), the solution $[U]_1^{16}$ reads
\begin{align}\label{Ugeo}
{[U]_1^{16}} =({{\cal D}_{33'2'1'\bar 3}} - 1)\left( {{\eta _{1'}} + {{\cal D}_{2'}}{\eta _{3'}} + {{\cal D}_{2'1'}}{\eta _{2'}} + {{\cal D}_{2'1'3'}}{\eta _1}} -{\eta _1}\right)
 - ({{\cal D}_{2'1'3'}} - 1)\left( {{{\cal D}_3}{\eta _{2'}} + {{\cal D}_{33'}}{\eta _{1'}} + {{\cal D}_{33'2'}}{\eta _{3'}}} \right).
\end{align}
One rewrites the relevant terms of the modified first-generation $U(t)$ combination Eq.~\eqref{U1},
we have 
\begin{subequations}
\begin{align}
{{\cal D}_3}\left( {{\eta _{2'}} + {{\cal D}_{3'}}{\eta _{1'}} + {{\cal D}_{3'2'}}{\eta _{3'}}} \right) = {\cal D}_3({\cal D}_{3'2'1'} - 1){p_2} = ({\cal D}_{33'2'1'\bar 3} - 1){\cal D}_3{p_2},\label{uu1}\\
{\eta _{1'}} + {{\cal D}_{2'}}{\eta _{3'}} + {{\cal D}_{2'1'}}{\eta _{2'}} + {{\cal D}_{2'1'3'}}{\eta _1} - {\eta _1} = ({{\cal D}_{2'1'3'}} - 1){{\cal D}_3}{p_2} ,\label{uu2}
\end{align}
\end{subequations}
By multiplying Eq.~\eqref{uu1} by $({{\cal D}_{2'1'3'}} - 1)$ and Eq.~\eqref{uu2} by $({\cal D}_{33'2'1'\bar 3} - 1)$, one rewrites $[U]_{stan}^{16}$ as
\begin{align}\label{Umathstan}
{[U]_{stan}^{16}} = ({{\cal D}_{33'2'1'\bar 3}} - 1)\left( {{\eta _{1'}} + {{\cal D}_{2'}}{\eta _{3'}} + {{\cal D}_{2'1'}}{\eta _{2'}} + {{\cal D}_{2'1'3'}}{\eta _1}} -{\eta _1}\right) 
 - ({{\cal D}_{2'1'3'}} - 1)\left( {{{\cal D}_3}{\eta _{2'}} + {{\cal D}_{33'}}{\eta _{1'}} + {{\cal D}_{33'2'}}{\eta _{3'}}} \right)
\end{align}
and therefore $[U]_1^{16}$ is indeed the second-generation Relay combination in the literature~\cite{tdi-otto-2015}.

\begin{figure}[!t]
\includegraphics[width=0.60\textwidth]{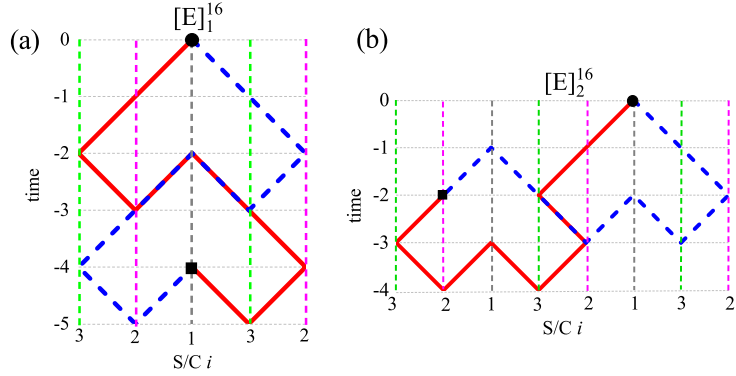}
\caption{\label{fig9} The space-time diagrams for the sixteen-link Monitor-type geometric TDI combinations.}
\end{figure}

Similarly, the combinations $[E]_1^{16}$ and $[E]_2^{16}$ shown in Fig.~\ref{fig9} are the Monitor-type TDI solutions.
The geometric TDI solution given in  Fig.~\ref{fig9}(a) implies
\begin{align}\label{Egeo}
[E]_1^{16} =({\cal D}_{2'1'1\bar 2'}-1)({\cal D}_{311'\bar 3}-1)\eta _1 -({\cal D}_{2'1'1\bar 2'}-1)({\cal D}_3{\eta _2} + {\cal D}_{31}\eta _{3'})
- ({\cal D}_{311'\bar 3}-1)({\cal D}_{2'1'1\bar 2'}-1)\eta _{1'} +({\cal D}_{311'\bar 3}-1) ({\cal D}_{2'}\eta _{3'} + {\cal D}_{2'1'}\eta _2) .
\end{align}
To compare with the combination $[E]_{stan}^{16}$ of the Monitor-type in the literature~\cite{tdi-d88}, one simplies individual terms of Eq.~\eqref{E1} as follows
\begin{subequations}
\begin{align}
\left(1 - {\cal D}_{311'\bar 3} \right)\eta_1+{\cal D}_3{\eta_2} + {\cal D}_{31}\eta _{3'}& =- \left( 1 - {\cal D}_{311'\bar 3} \right)p_1,\label{ee1}\\
\left( 1 - {\cal D}_{2'1'1\bar 2'} \right)\eta _{1'}+\left( {\cal D}_{2'}\eta_{3'} + {\cal D}_{2'1'}\eta _2 \right) & = -\left( 1 - {\cal D}_{2'1'1\bar 2'} \right)p_1,\label{ee2}
\end{align}
\end{subequations}
By multiplying Eq.~\eqref{ee1}by $( 1 - {\cal D}_{2'1'1\bar 2'})$ and Eq.~\eqref{ee2} by $(1 - {\cal D}_{311'\bar 3})$, one rewrites $[E]_{stan}^{16}$ as
\begin{align}\label{Emath}
[ E ]_{stan}^{16} = \left( {1 - {{\cal D}_{2'1'1\bar 2'}}} \right)\left[ {\left( {1 - {{\cal D}_{311'\bar 3}}} \right){\eta _1} + {{\cal D}_3}{\eta _2} + {{\cal D}_{31}}{\eta _{3'}}} \right]
 - \left( {1 - {{\cal D}_{311'\bar 3}}} \right)\left[ {\left( {1 - {{\cal D}_{2'1'1\bar 2'}}} \right){\eta _{1'}}- {{\cal D}_{2'}}{\eta _{3'}} - {{\cal D}_{2'1'}}{\eta _2}} \right] .
\end{align}
The above equation is readily compared to Eq.~\eqref{Egeo}, and therefore $[E]_1^{16}$ is indeed the second-generation Monitor combination in the literature~\cite{tdi-otto-2015}.

\begin{figure}[!t]
\includegraphics[width=0.60\textwidth]{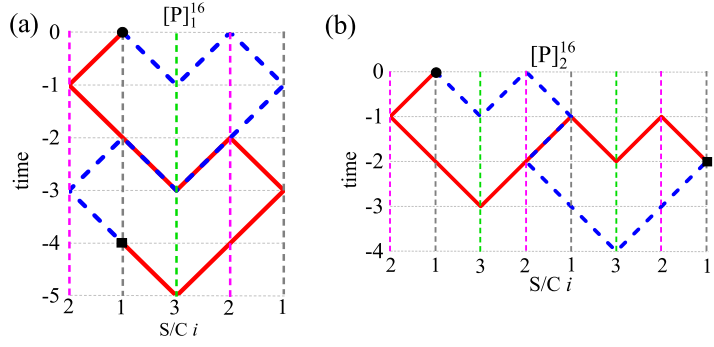}
\caption{\label{fig10} The space-time diagrams for the sixteen-link Beacon-type geometric TDI combinations.}
\end{figure}

The space-time diagram of the Beacon-type TDI combination $[P]_1^{16}, [P]_2^{16}$ are illustrated in Fig.~\ref{fig10}.
Similar to the above analysis, one of the geometric TDI solutions, $[P]_1^{16}$, is shown to be equivalent to the standard form in the literature~\cite{tdi-otto-2015}.

Moreover, it is interesting that the modified second-generation TDI combinations can be derived from the modified first-generation counterpart with appropriate time shifts.
This can be seen by comparing the relevant structures of the respective space-time diagrams.
To be specific, one has the following relations.
For Michelson-type TDI combination $[X]_1^{16}$ and $[X]_2^{16}$, we have
\begin{align}\label{MICH}
[X]_1^{16}(t) \approx & X(t) - X(t -4T),\\\notag
[X]_2^{16}(t) \approx & X(t) - X(t -3T).
\end{align}
For Relay-type TDI combination $[U]_1^{16},[U]_2^{16}$ and $[U]_3^{16}$,
\begin{align}\label{RELAY}
[U]_1^{16}(t) \approx & U(t) - U(t-3T),\\\notag
[U]_2^{16}(t) \approx & U(t) - U(t-2T),\\\notag
[U]_3^{16}(t) \approx & U(t) - U(t-T).
\end{align}
For Monitor-type TDI combination $[E]_1^{16}$ and $[E]_2^{16}$,
\begin{align}\label{MONI}
[E]_1^{16}(t) \approx & E(t) - E(t-2T),\\\notag
[E]_2^{16}(t) \approx & E(t) - E(t-T).
\end{align}
For Beacon-type TDI combination $[P]_1^{16}$ and $[P]_2^{16}$,
\begin{align}\label{BEAN}
[P]_1^{16}(t) \approx & P(t) - P(t-2T),\\\notag
[P]_2^{16}(t) \approx & P(t) - P(t-T).
\end{align}

Besides the TDI that have already explored in the literature~\cite{tdi-d88}, it is worth noting that the other combinations, namely, $[X]_2^{16},[U]_2^{16},[U]_3^{16},[E]_2^{16},[P]_2^{16}$, are largely new results which will be referred to as {\it alternative} solutions. 
The remainder of the present paper is mainly devoted to studying their properties.
First, these alternative solutions are featured by a reduced temporal span in the space-time diagram.
As a result, the resultant contamination of the data set is likely to be minimized in the presence of instrumental gaps or glitches, as pointed out in~\cite{tdi-geometric-2005}.

We give in the second and third columns of Tab.~\ref{16linkTab}, the coefficients $d_i, d_{i'}$ and $f_i$ defined earlier by Eq.~\eqref{delat} of the sixteen-link modified second-generation geometric TDI combinations.
These quantities are then utilized to eventually evaluate the residual $\frac{c\delta t}{L}$ and shown by the fourth column of the table (see the derivations given in Sec.~\ref{section4}).

\begin{table}
\caption{Delay-time residual of the sixteen-link modified second-generation geometric TDI combinations. 
The fifth and sixth columns were evaluated using the parameters of the LISA detector.}
\centering
\newcommand{\tabincell}[2]{\begin{tabular}{@{}#1@{}}#2\end{tabular}}
\renewcommand\arraystretch{2}
\begin{tabular}{|c|c|c|c|c|}
 \hline
 \hline
\makecell[c] {TDI\\ combination} & $\{d_i,d_{i'}\}$ &$ \{f_i\}$ &time residual $\frac{c\delta t}{L}$&\makecell[c] {coefficient\\ before $\cos\Omega t$ (s)}\\
 \hline
 $[X]_1^{16}$ & $\{0,0,0,0,0,0\}$& $\{0,-16,16\}$&$16\frac{L}{c^2}(\ddot L_3-\ddot L_2)$&$7.2\times10^{-13}$\\
 \hline
 $[X]_2^{16}$ & $\{0,0,0,0,0,0\}$& $\{0,-8,8\}$&$8\frac{L}{c^2}(\ddot L_3-\ddot L_2)$&$3.6\times10^{-13}$\\
 \hline
 $[U]_1^{16}$ &$\{0,0,0,0,0,0\}$& $\{-6,-6,12\}$&$6\frac{L}{c^2}(2\ddot L_3-\ddot L_2-\ddot L_1)$&$2.7\times10^{-13}$\\
 \hline
 $[U]_2^{16}$ &$\{0,0,0,0,0,0\}$& $\{-4,-4,8\}$&$4\frac{L}{c^2}(2\ddot L_3-\ddot L_2-\ddot L_1)$&$1.8\times10^{-13}$\\
 \hline
 $[U]_3^{16}$ &$\{0,0,0,0,0,0\}$& $\{-2,-2,4\}$&$4\frac{L}{c^2}(2\ddot L_3-\ddot L_2-\ddot L_1)$&$9\times10^{-14}$\\
 \hline
 $[E]_1^{16}$ &$\{0,0,0,0,0,0\}$& $\{0,-4,4\}$&$4\frac{L}{c^2}(\ddot L_3-\ddot L_2)$&$1.8\times10^{-13}$\\
 \hline
 $[E]_2^{16}$ &$\{0,0,0,0,0,0\}$& $\{0,-2,2\}$&$2\frac{L}{c^2}(\ddot L_3-\ddot L_2)$&$9\times10^{-14}$\\
 \hline
 $[P]_1^{16}$ &$\{0,0,0,0,0,0\}$& $\{4,-4,0\}$&$4\frac{L}{c^2}(\ddot L_1-\ddot L_2)$&$1.8\times10^{-13}$\\
 \hline
 $[P]_2^{16}$ &$\{0,0,0,0,0,0\}$& $\{2,-2,0\}$&$2\frac{L}{c^2}(\ddot L_1-\ddot L_2)$&$9\times10^{-14}$\\
 \hline
 \hline
\end{tabular}
\label{16linkTab}
\end{table}

For the sixteen-link modified second-generation TDI combinations, the coefficients $d_i$ and $d_{i'}$ vanish identically, which implies that residuals of the virtual optical paths only come from the armlength acceleration terms.
As will be specified below, for the sixteen-link second-generation TDI combinations, there are a total of thirty-one solutions, from which thirteen of them (given in Tab.~\ref{16linkTraje2}) can be intuitively derived using their modified first-generation counterparts. 
In this context, the remaining eighteen combinations (given in Tab.~\ref{16linkTra3}) are generic.
 
\begin{table}
\caption{Trajectory of the second-generation TDI combination laser link. These thirteen second-generation TDI combination solutions can be intuitively derived using their modified first-generation couterparts. }
\centering
\newcommand{\tabincell}[2]{\begin{tabular}{@{}#1@{}}#2\end{tabular}}
\renewcommand\arraystretch{2}
\begin{tabular}{|c|c|}
\hline
\hline
TDI combination & Laser link trajectory \\
 \hline
$[U]_4^{16}$&$1 \leftarrow 2 \leftarrow 1 \leftarrow 3 \leftarrow 2 \leftarrow 3 \leftarrow 1 \leftarrow 2 \leftarrow 1 \to 3 \to 2 \to 1 \to 2 \to 1 \to 2 \to 3 \to 1$\\
\hline
$[U]_5^{16}$&$1 \leftarrow2 \leftarrow 1 \leftarrow 3 \leftarrow 2 \leftarrow 1 \to 2 \leftarrow 3 \leftarrow 1 \leftarrow 2 \leftarrow 1 \to 3 \to 2 \to 1 \to 2 \to 3 \to 1$\\
\hline
$[U]_6^{16}$&$1 \leftarrow 2 \leftarrow 1 \leftarrow 3 \leftarrow 1 \leftarrow 2 \leftarrow 1 \to 3 \to 2 \to 1 \to 2 \leftarrow 3 \leftarrow 2 \to 1 \to 2 \to 3 \to 1$\\
\hline
$[PE]_1^{16}$&$1 \leftarrow 2 \leftarrow 3 \leftarrow 2 \leftarrow 1 \to 3 \to 2 \to 3 \leftarrow 1 \to 2 \leftarrow 3 \leftarrow 2 \to 1 \leftarrow 3 \to 2 \to 3 \to 1$\\
\hline
$[PE]_2^{16}$&$1 \leftarrow 2 \leftarrow 3 \leftarrow 2 \to 1 \to 2 \leftarrow 3 \leftarrow 2 \leftarrow 1 \to 3 \to 2 \to 3 \leftarrow 1 \leftarrow 3 \to 2 \to 3 \to 1$\\
\hline
$[PE]_3^{16}$&$1 \leftarrow 2 \leftarrow 3 \leftarrow 2 \leftarrow 3 \leftarrow 2 \leftarrow 1 \leftarrow 3 \to 2 \to 3 \leftarrow 1 \to 2 \to 1 \leftarrow 3 \to 2 \to 3 \to 1$\\
\hline
$[PE]_4^{16}$&$1 \leftarrow 2 \leftarrow 1 \leftarrow 3 \leftarrow 1 \leftarrow 2 \leftarrow 1 \to 3 \leftarrow 2 \to 1 \to 2 \to 3 \to 2 \to 1 \to 2 \leftarrow 3 \to 1$\\
\hline
$[PE]_5^{16}$&$1 \leftarrow 2 \leftarrow 1 \leftarrow 3 \leftarrow 2 \to 1 \to 2 \to 1 \to 2 \leftarrow 3 \leftarrow 1 \leftarrow 2 \leftarrow 1 \to 3 \to 2 \to 3 \to 1$\\
\hline
$[PE]_6^{16}$&$1 \leftarrow 2 \leftarrow 1 \leftarrow 3 \leftarrow 2 \to 1 \to 2 \to 3 \leftarrow 1 \leftarrow 2 \leftarrow 1 \to 3 \to 2 \to 1 \to 2 \leftarrow 3 \to 1$\\
\hline
$[PE]_7^{16}$&$1 \leftarrow 2 \leftarrow 1 \leftarrow 3 \to 2 \to 1\to 2 \leftarrow 3 \leftarrow 1 \leftarrow 2 \leftarrow 1 \to 3 \leftarrow 2 \to 1 \to 2 \to 3 \to 1$\\
\hline
$[PE]_8^{16}$&$1 \leftarrow 2 \leftarrow 1 \leftarrow 3 \to 2 \to 3 \leftarrow 1 \leftarrow 2 \leftarrow 1 \to 3 \leftarrow 2 \to 1 \to 2 \to 1 \to 2 \leftarrow 3 \to 1$\\
\hline
$[PE]_9^{16}$&$1 \leftarrow 2 \leftarrow 1 \leftarrow 2 \leftarrow 1 \to 3 \to 2 \to 3 \leftarrow 1 \leftarrow 3 \leftarrow 2 \to 1 \to 2 \to 1 \to 2 \leftarrow 3 \to 1$\\
\hline
$[PE]_{10}^{16}$&$1 \leftarrow 2 \leftarrow 1 \leftarrow 2 \leftarrow 1 \to 3 \leftarrow 2 \to 1 \to 2 \to 1 \to 2 \leftarrow 3 \leftarrow 1 \leftarrow 3 \to 2 \to 3 \to 1$\\
 \hline
 \hline
\end{tabular}
\label{16linkTraje2}
\end{table}

\begin{figure}[!t]
\includegraphics[width=0.60\textwidth]{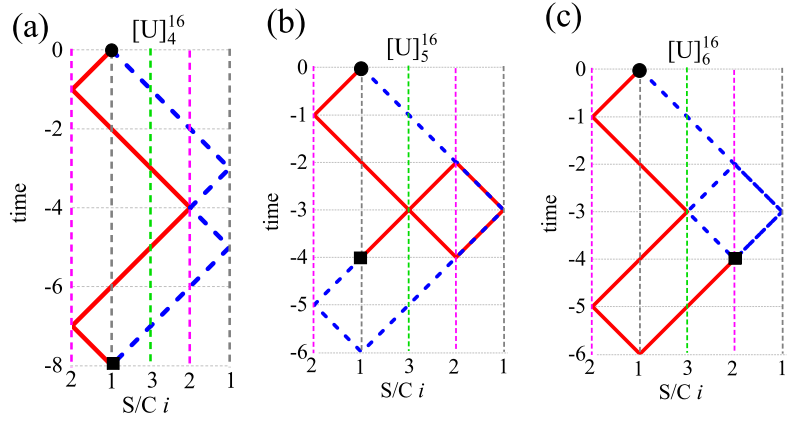}
\caption{\label{fig11} The space-time diagrams for the sixteen-link Relay-type second-generation geometric TDI combinations.}
\end{figure}

The combinations $[U]_4^{16}$, $[U]_5^{16}$, and $[U]_6^{16}$ shown in Fig.~\ref{fig11} can be rewritten, up to second-order commutators, using the modified first generation Relay combinations, as follows: 
\begin{align}\label{16linkuapp}
[U]_4^{16}(t) &\simeq U(t) + {\bar U}(t - 4T),\\\notag
[U]_5^{16}(t)& \simeq [U]_6^{16}(t) \simeq  U(t) + {\bar U}(t - 2T).
\end{align}
where ${\bar U}$ denotes the flipped $U$ combination.
Besides Eq.~\eqref{16linkuapp}, the relevant grids in the space-time diagram for $[U]_5^{16}$ and $[U]_6^{16}$ are the same, but the specific links are distinct, as given by Tab.~\ref{16linkTraje2}.

\begin{figure}[!t]
\includegraphics[width=0.60\textwidth]{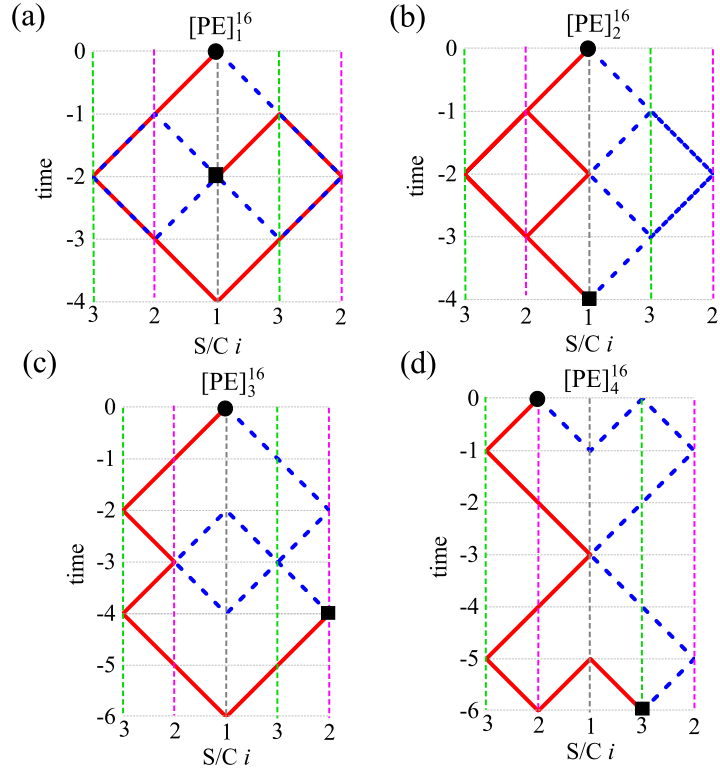}
\caption{\label{fig12}The space-time diagrams for the sixteen-link Beacon and Monitor combination-type second-generation geometric TDI combinations.}
\end{figure}

The combination $[PE]_1^{16}$, $[PE]_2^{16}$, $[PE]_3^{16}$, and $[PE]_4^{16}$ shown in Fig.~\ref{fig12} can be furnished by the modified first-generation Beacon and modified Monitor combination as follows:
\begin{align}\label{16linkpeapp}
[PE]_1^{16}(t) \simeq & [PE]_2^{16}(t) \simeq E(t) + P(t-T),\\\notag
[PE]_3^{16}(t) \simeq & E(t) + P(t-3T),\\\notag
[PE]_4^{16}(t) \simeq & P(t)+E(t-3T).
\end{align}

\begin{figure}[!t]
\includegraphics[width=0.60\textwidth]{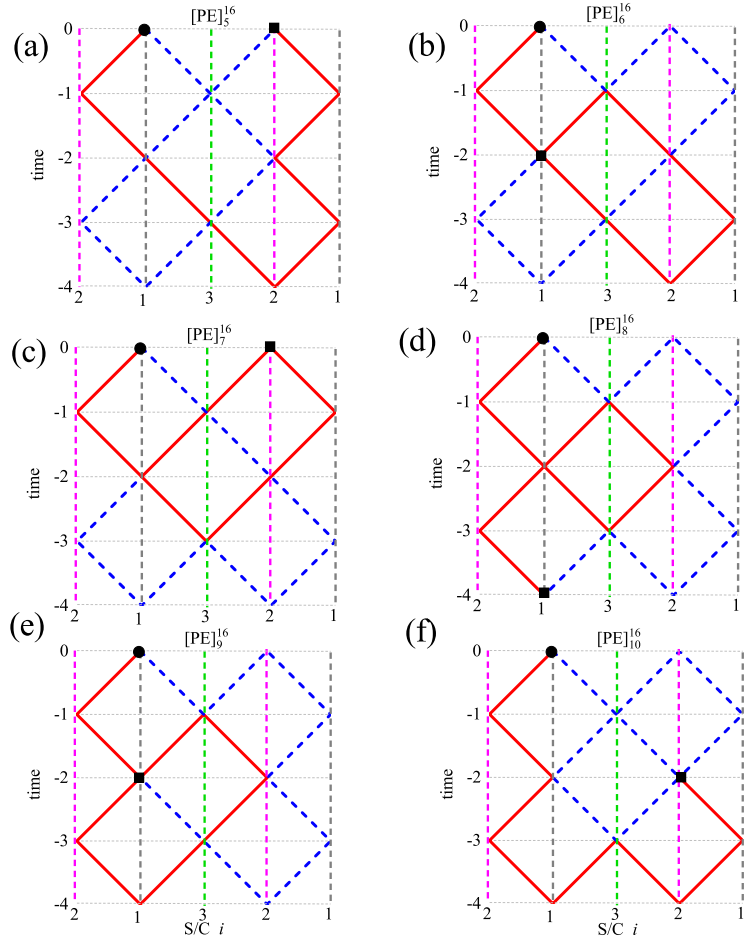}
\caption{\label{fig13} The space-time diagrams for the sixteen-link Relay-type or Beacon and Monitor combination type second-generation geometric TDI combinations.}
\end{figure}

In addition, $[PE]_5^{16}-[PE]_{10}^{16}$ are illustrated in Fig.~\ref{fig13}, it consists of the Beacon and Monitor combination, or two relay combination:
\begin{align}\label{16linkpeuu}
[PE]_5^{16}(t) \simeq [PE]_6^{16}(t) \simeq [PE]_7^{16}(t)\simeq [PE]_8^{16}(t) \simeq [PE]_9^{16}(t) \simeq [PE]_{10}^{16}(t)\simeq P(t) + E(t - T) \simeq U(t)+{\bar U}(t).
\end{align}
Although the relevant grids are identical, the links for the diagrams  $[PE]_5^{16}$ to $[PE]_{10}^{16}$ are different.

\begin{table}
\caption{Delay-time residual of the sixteen-link second generation geometric TDI combinations.These thirteen second-generation TDI combination solutions can be intuitively derived using their modified first-generation couterparts. 
The fifth and sixth columns were evaluated using the parameters of the LISA detector.}
\centering
\newcommand{\tabincell}[2]{\begin{tabular}{@{}#1@{}}#2\end{tabular}}
\renewcommand\arraystretch{2}
\begin{tabular}{|c|c|c|c|c|c|}
 \hline
 \hline
\makecell[c] {TDI\\ combin-\\ation} & $\{d_i,d_{i'}\}$ &$ \{f_i\}$ &time residual $\frac{c\delta t}{L}$&\makecell[c] {coefficient\\ before $\sin3\Omega t$ (s)}&\makecell[c] {coefficient\\ before $\cos\Omega t$ (s)}\\
 \hline
 $[U]_4^{16}$ &$\{2,2,2,-2,-2,-2\}$& $\{-6,-10,16\}$&\makecell[c]{$2\sum\limits_{i = 1}^3 \frac{\dot{L}_i - \dot{L}_{i'}}{c}$+\\$2\frac{L}{c^2}(8\ddot L_3-5\ddot L_2-3\ddot L_1)$}&$1.8\times10^{-13}$&$4.5\times10^{-13}$\\
 \hline
 $[U]_5^{16}$ &$\{1,1,1,-1,-1,-1\}$& $\{-2,-6,8\}$&\makecell[c]{$\sum\limits_{i = 1}^3 \frac{\dot{L}_i - \dot{L}_{i'}}{c}$+\\$2\frac{L}{c^2}(4\ddot L_3-3\ddot L_2-\ddot L_1)$}&$9\times10^{-14}$&$2.7\times10^{-13}$\\
 \hline
 $[U]_6^{16}$ &$\{4,4,4,-4,-4,-4\}$& $\{-2,-6,8\}$&\makecell[c]{$4\sum\limits_{i = 1}^3 \frac{\dot{L}_i - \dot{L}_{i'}}{c}$+\\$2\frac{L}{c^2}(4\ddot L_3-3\ddot L_2-\ddot L_1)$}&$3.6\times10^{-13}$&$2.7\times10^{-14}$\\
 \hline
 $[PE]_1^{16}$ &$\{1,1,1,-1,-1,-1\}$& $\{0,-2,2\}$&$\sum\limits_{i = 1}^3 \frac{\dot{L}_i - \dot{L}_{i'}}{c}+2\frac{L}{c^2}(\ddot L_3-\ddot L_2)$&$9\times10^{-14}$&$9\times10^{-14}$\\
 \hline
 $[PE]_2^{16}$ &$\{3,3,3,-3,-3,-3\}$& $\{0,-2,2\}$&$3\sum\limits_{i = 1}^3 \frac{\dot{L}_i - \dot{L}_{i'}}{c}2+\frac{L}{c^2}(\ddot L_3-\ddot L_2)$&$2.7\times10^{-14}$&$9\times10^{-14}$\\
 \hline
 $[PE]_3^{16}$ &$\{4,4,4,-4,-4,-4\}$& $\{0,-6,6\}$&$4\sum\limits_{i = 1}^3 \frac{\dot{L}_i - \dot{L}_{i'}}{c}+6\frac{L}{c^2}(\ddot L_3-\ddot L_2)$&$3.6\times10^{-14}$&$2.7\times10^{-13}$\\
 \hline
 $[PE]_4^{16}$ &$\{5,5,5,-5,-5,-5\}$& $\{6,-6,0\}$&$5\sum\limits_{i = 1}^3 \frac{\dot{L}_i - \dot{L}_{i'}}{c}+6\frac{L}{c^2}(\ddot L_1-\ddot L_2)$&$4.5\times10^{-14}$&$2.7\times10^{-13}$\\
 \hline
 $[PE]_5^{16}$ &$\{1,1,1,-1,-1,-1\}$& $\{2,-2,0\}$&$\sum\limits_{i = 1}^3 \frac{\dot{L}_i - \dot{L}_{i'}}{c}+2\frac{L}{c^2}(\ddot L_1-\ddot L_2)$&$9\times10^{-14}$&$9\times10^{-14}$\\
 \hline
 $[PE]_6^{16}$ &$\{-1,-1,-1,1,1,1\}$& $\{2,-2,0\}$&$-\sum\limits_{i = 1}^3 \frac{\dot{L}_i - \dot{L}_{i'}}{c}+2\frac{L}{c^2}(\ddot L_1-\ddot L_2)$&$9\times10^{-14}$&$9\times10^{-14}$\\
 \hline
 $[PE]_7^{16}$ &$\{1,1,1,-1,-1,-1\}$& $\{2,-2,0\}$&$\sum\limits_{i = 1}^3 \frac{\dot{L}_i - \dot{L}_{i'}}{c}+2\frac{L}{c^2}(\ddot L_1-\ddot L_2)$&$9\times10^{-14}$&$9\times10^{-14}$\\
 \hline
 $[PE]_8^{16}$ &$\{1,1,1,-1,-1,-1\}$& $\{2,-2,0\}$&$\sum\limits_{i = 1}^3 \frac{\dot{L}_i - \dot{L}_{i'}}{c}+2\frac{L}{c^2}(\ddot L_1-\ddot L_2)$&$9\times10^{-14}$&$9\times10^{-14}$\\
 \hline
 $[PE]_9^{16}$ &$\{3,3,3,-3,-3,-3\}$& $\{2,-2,0\}$&$3\sum\limits_{i = 1}^3 \frac{\dot{L}_i - \dot{L}_{i'}}{c}+2\frac{L}{c^2}(\ddot L_1-\ddot L_2)$&$2.7\times10^{-14}$&$9\times10^{-14}$\\
 \hline
 $[PE]_{10}^{16}$ &$\{-1,-1,-1,1,1,1 \}$& $\{2,-2,0\}$&$-\sum\limits_{i = 1}^3 \frac{\dot{L}_i - \dot{L}_{i'}}{c}+2\frac{L}{c^2}(\ddot L_1-\ddot L_2)$&$9\times10^{-14}$&$9\times10^{-14}$\\
 \hline
 \hline
\end{tabular}
\label{16linkTab11}
\end{table}
We give in the second and third columns of Tab.~\ref{16linkTab11} and Tab.~\ref{16linkTab22}, the coefficients $d_i, d_{i'}$ and $f_i$ defined earlier by Eq.~\eqref{delat} of the sixteen-link second-generation geometric TDI combinations.
These quantities are then utilized to eventually evaluate the residual $\frac{c\delta t}{L}$ and shown by the fourth column of the table (see the derivations given in Sec.~\ref{section4}).

\begin{table}
\caption{Trajectory of the second-generation TDI combination laser link. 
These eighteen second-generation combinations are generic ones that do not belong to any existing class.}
\centering
\newcommand{\tabincell}[2]{\begin{tabular}{@{}#1@{}}#2\end{tabular}}
\renewcommand\arraystretch{2}
\begin{tabular}{|c|c|}
 \hline
 \hline
TDI combination & Laser link trajectory \\
 \hline
 $[T]_1^{16}$&$1 \leftarrow 2 \leftarrow 1 \leftarrow 3 \to 2 \to 1 \leftarrow 3 \leftarrow 2 \leftarrow 1 \to 3 \to 1 \to 2 \leftarrow 3 \leftarrow 1 \to 2 \to 3 \to 1$\\
 \hline
$[T]_2^{16}$&$1 \leftarrow 2 \leftarrow 1 \leftarrow 2 \leftarrow 1 \to 3 \to 2 \to 1 \leftarrow 3 \leftarrow 2 \to 1 \to 2 \leftarrow 3 \leftarrow 1 \to 2 \to 3 \to 1$\\
 \hline
 $[T]_3^{16}$&$1 \leftarrow 2 \leftarrow 3 \leftarrow 2 \leftarrow 1 \leftarrow 3 \leftarrow 2 \leftarrow 1 \to 3 \to 2 \to 1 \to 2 \to 3 \leftarrow 1 \to 2 \to 3 \to 1$\\
 \hline
 $[T]_4^{16}$&$1 \leftarrow 2 \leftarrow 3 \leftarrow 2 \leftarrow 1 \to 3 \leftarrow 2 \to 1 \to 2 \to 1 \leftarrow 3 \leftarrow 1 \leftarrow 2 \to 3 \to 2 \to 3 \to 1$\\
 \hline
 $[T]_5^{16}$&$1 \leftarrow 2 \leftarrow 3 \leftarrow 2 \leftarrow 3 \leftarrow 2 \leftarrow 1 \to 3 \to 2 \to 1 \leftarrow 3 \to 2 \to 3 \leftarrow 1 \to 2 \to 3 \to 1$\\
\hline
 $[T]_6^{16}$&$1 \leftarrow 2 \leftarrow 3 \leftarrow 1 \leftarrow 3 \leftarrow 2 \to 1 \leftarrow 3 \leftarrow 2 \leftarrow 1 \to 3 \to 2 \to 3 \to 1 \to 2 \to 3 \to 1$\\
 \hline
 $[T]_7^{16}$&$1 \leftarrow 2 \leftarrow 3 \leftarrow 1 \leftarrow 3 \leftarrow 1 \leftarrow 3 \leftarrow 2 \leftarrow 1 \to 3 \to 2 \to 1 \to 3 \to 1 \to 2 \to 3 \to 1$\\
  \hline
 $[T]_8^{16}$&$1 \leftarrow 2 \leftarrow 3 \leftarrow 1 \leftarrow 3 \to 2 \leftarrow 1 \to 3 \to 2 \to 1 \leftarrow 3 \leftarrow 2 \leftarrow 3 \to 1 \to 2 \to 3 \to 1$\\
 \hline
  $[T]_9^{16}$&$1 \leftarrow 2 \leftarrow 1 \leftarrow 2 \to 3 \to 2 \to 1 \leftarrow 3 \leftarrow 2 \to 1 \leftarrow 3 \leftarrow 1 \to 2 \leftarrow 3 \to 1 \to 3 \to 1$\\
  \hline
   $[T]_{10}^{16}$&$1 \leftarrow 2 \leftarrow 3 \leftarrow 2 \to 1 \to 2 \to 1 \leftarrow 3 \leftarrow 1 \leftarrow 3 \to 2 \leftarrow 1 \to 3 \to 1 \leftarrow 2 \to 3 \to 1$\\
  \hline
   $[T]_{11}^{16}$&$1 \leftarrow 2 \leftarrow 1 \leftarrow 3 \leftarrow 2 \to 1 \to 3 \to 2 \to 1 \leftarrow 3 \leftarrow 1 \leftarrow 2 \to 3 \to 1 \to 2 \leftarrow 3 \to 1$\\
  \hline
   $[T]_{12}^{16}$&$1 \leftarrow 2 \leftarrow 1 \leftarrow 3 \leftarrow 2 \to 1 \leftarrow 3 \to 2 \leftarrow 1 \to 3 \to 1 \to 2 \to 3 \leftarrow 1 \to 2 \leftarrow 3 \to 1$\\
   \hline
   $[T]_{13}^{16}$&$1 \leftarrow 2\leftarrow 1 \leftarrow 3 \to 2 \to 1 \to 3 \leftarrow 2 \to 1 \leftarrow 3 \leftarrow 1 \leftarrow 2 \leftarrow 3 \to 1 \to 2 \to 3 \to 1$\\
   \hline
   $[T]_{14}^{16}$&$1 \leftarrow 2 \leftarrow 1 \leftarrow 3 \to 2 \to 1 \leftarrow 3 \to 2 \to 3 \leftarrow 1 \to 2 \leftarrow 3 \leftarrow 2 \leftarrow 3 \to 1 \to 3 \to 1$\\
   \hline
   $[T]_{15}^{16}$&$1 \leftarrow 2 \leftarrow 1 \leftarrow 2 \leftarrow 1 \to 3 \leftarrow 2 \to 1 \leftarrow 3 \to 2 \to 1 \to 2 \to 3 \leftarrow 1 \to 2 \leftarrow 3 \to 1$\\
   \hline
   $[T]_{16}^{16}$&$1 \leftarrow 2 \leftarrow 1 \leftarrow 2 \leftarrow 3 \to 1 \to 2 \to 1 \to 3 \leftarrow 2 \to 1 \leftarrow 3 \leftarrow 1 \leftarrow 3 \to 2 \to 3 \to 1$\\
   \hline
   $[T]_{17}^{16}$&$1 \leftarrow 2 \leftarrow 1 \leftarrow 2 \to 3 \to 2 \to 1 \leftarrow 3 \leftarrow 1 \leftarrow 3 \leftarrow 2 \to 1 \to 3 \to 1 \to 2 \leftarrow 3 \to 1$\\
   \hline
   $[T]_{18}^{16}$&$1 \leftarrow 2 \leftarrow 1 \to 3 \leftarrow 2 \to 1 \to 2 \leftarrow 3 \leftarrow 2 \to 1 \leftarrow 3 \to 2 \to 3 \leftarrow 1 \leftarrow 2 \to 3 \to 1$\\
 \hline
 \hline
\end{tabular}
\label{16linkTra3}
\end{table}

\newpage

\begin{table}
\caption{Delay-time residual of the sixteen-link second-generation geometric TDI combinations. 
These eighteen second-generation combinations are generic ones that do not belong to any existing class.
The fifth and sixth columns were evaluated using the parameters of the LISA detector.}
\centering
\newcommand{\tabincell}[2]{\begin{tabular}{@{}#1@{}}#2\end{tabular}}
\renewcommand\arraystretch{2}
\begin{tabular}{|c|c|c|c|c|c|}
 \hline
 \hline
\makecell[c] {TDI\\ combination} & $\{d_i,d_{i'}\}$ &$ \{f_i\}$ &time residual $\frac{c\delta t}{L}$&\makecell[c] {coefficient\\ before $\sin3\Omega t$ (s)}&\makecell[c] {coefficient\\ before $\cos\Omega t$ (s)}\\
 \hline
 $[T]_1^{16}$ &$\{1,1,1,-1,-1,-1\}$& $\{0,-2,2\}$&$\sum\limits_{i = 1}^3 \frac{\dot{L}_i - \dot{L}_{i'}}{c}+2\frac{L}{c^2}(\ddot L_3-\ddot L_2)$&$9\times10^{-14}$&$9\times10^{-14}$\\
 \hline
 $[T]_2^{16}$ &$\{1,1,1,-1,-1,-1\}$& $\{0,-2,2\}$&$\sum\limits_{i = 1}^3 \frac{\dot{L}_i - \dot{L}_{i'}}{c}+2\frac{L}{c^2}(\ddot L_3-\ddot L_2)$&$9\times10^{-14}$&$9\times10^{-14}$\\
 \hline
 $[T]_3^{16}$ &$\{4,4,4,-4,-4,-4\}$& $\{0,-8,8\}$&$4\sum\limits_{i = 1}^3 \frac{\dot{L}_i - \dot{L}_{i'}}{c}+8\frac{L}{c^2}(\ddot L_3-\ddot L_2)$&$3.6\times10^{-13}$&$3.6\times10^{-13}$\\
 \hline
 $[T]_4^{16}$ &$\{1,1,1,-1,-1,-1\}$& $\{0,-2,2\}$&$\sum\limits_{i = 1}^3 \frac{\dot{L}_i - \dot{L}_{i'}}{c}+2\frac{L}{c^2}(\ddot L_3-\ddot L_2)$&$9\times10^{-13}$&$9\times10^{-14}$\\
 \hline
 $[T]_5^{16}$ &$\{3,3,3,-3,-3,-3\}$& $\{0,-6,6\}$&$3\sum\limits_{i = 1}^3 \frac{\dot{L}_i - \dot{L}_{i'}}{c}+6\frac{L}{c^2}(\ddot L_3-\ddot L_2)$&$2.7\times10^{-13}$&$2.7\times10^{-13}$\\
 \hline
 $[T]_6^{16}$ &$\{4,4,4,-4,-4,-4\}$& $\{0,-8,8\}$&$4\sum\limits_{i = 1}^3 \frac{\dot{L}_i - \dot{L}_{i'}}{c}+8\frac{L}{c^2}(\ddot L_3-\ddot L_2)$&$3.6\times10^{-13}$&$3.6\times10^{-13}$\\
 \hline
 $[T]_7^{16}$ &$\{5,5,5,-5,-5,-5\}$& $\{0,-10,10\}$&$5\sum\limits_{i = 1}^3 \frac{\dot{L}_i - \dot{L}_{i'}}{c}+10\frac{L}{c^2}(\ddot L_3-\ddot L_2)$&$4.5\times10^{-13}$&$4.5\times10^{-13}$\\
 \hline
 $[T]_8^{16}$ &$\{1,1,1,-1,-1,-1\}$& $\{0,-2,2\}$&$\sum\limits_{i = 1}^3 \frac{\dot{L}_i - \dot{L}_{i'}}{c}+2\frac{L}{c^2}(\ddot L_3-\ddot L_2)$&$9\times10^{-14}$&$9\times10^{-14}$\\
 \hline
 $[T]_9^{16}$ &$\{-1,-1,-1,1,1,1\}$& $\{0,0,0\}$&$-\sum\limits_{i = 1}^3 \frac{\dot{L}_i - \dot{L}_{i'}}{c}$&$9\times10^{-14}$&$0$\\
 \hline
 $[T]_{10}^{16}$ &$\{1,1,1,-1,-1,-1\}$& $\{0,0,0\}$&$\sum\limits_{i = 1}^3 \frac{\dot{L}_i - \dot{L}_{i'}}{c}$&$9\times10^{-14}$&$0$\\
 \hline
 $[T]_{11}^{16}$ &$\{1,1,1,-1,-1,-1\}$& $\{2,0,-2\}$&$\sum\limits_{i = 1}^3 \frac{\dot{L}_i - \dot{L}_{i'}}{c}+2\frac{L}{c^2}(\ddot L_1-\ddot L_3)$&$9\times10^{-14}$&$9\times10^{-14}$\\
 \hline
 $[T]_{12}^{16}$ &$\{3,3,3,-3,-3,-3\}$& $\{0,0,0\}$&$3\sum\limits_{i = 1}^3 \frac{\dot{L}_i - \dot{L}_{i'}}{c}$&$2.7\times10^{-13}$&$0$\\
 \hline
 $[T]_{13}^{16}$ &$\{-1,-1,-1,1,1,1 \}$& $\{2,-2,0\}$&$-\sum\limits_{i = 1}^3 \frac{\dot{L}_i - \dot{L}_{i'}}{c}+2\frac{L}{c^2}(\ddot L_1-\ddot L_2)$&$9\times10^{-14}$&$9\times10^{-14}$\\
 \hline
 $[T]_{14}^{16}$ &$\{1,1,1,-1,-1,-1\}$& $\{0,0,0\}$&$\sum\limits_{i = 1}^3 \frac{\dot{L}_i - \dot{L}_{i'}}{c}$&$9\times10^{-14}$&$0$\\
 \hline
 $[T]_{15}^{16}$ &$\{3,3,3,-3,-3,-3\}$& $\{0,0,0\}$&$3\sum\limits_{i = 1}^3 \frac{\dot{L}_i - \dot{L}_{i'}}{c}$&$2.7\times10^{-13}$&$0$\\
  \hline
 $[T]_{16}^{16}$ &$\{-1,-1,-1,1,1,1\}$& $\{2,-2,0\}$&$-\sum\limits_{i = 1}^3 \frac{\dot{L}_i - \dot{L}_{i'}}{c}+2\frac{L}{c^2}(\ddot L_1-\ddot L_2)$&$9\times10^{-14}$&$9\times10^{-14}$\\
 \hline
 $[T]_{17}^{16}$ &$\{1,1,1,-1,-1,-1\}$& $\{2,0,-2\}$&$-\sum\limits_{i = 1}^3 \frac{\dot{L}_i - \dot{L}_{i'}}{c}+2\frac{L}{c^2}(\ddot L_1-\ddot L_3)$&$9\times10^{-14}$&$9\times10^{-14}$\\
 \hline
 $[T]_{18}^{16}$ &$\{1,1,1,-1,-1,-1\}$& $\{0,0,0\}$&$\sum\limits_{i = 1}^3 \frac{\dot{L}_i - \dot{L}_{i'}}{c}$&$9\times10^{-14}$&$0$\\
 \hline
 \hline
\end{tabular}
\label{16linkTab22}
\end{table}
For the sixteen-link second-generation TDI combinations, the coefficients $d_i=  - d_{i'} \ne 0$, which implies that residuals of the virtual optical paths only come from the armlength velocity terms and armlength acceleration terms.

\section{Analytical results of delay-time residual}\label{section4}

In this section, we first derive the leading-order delay-time residuals analytically due to the optical path mismatch.
Physically, such an additional discrepancy originated from the fact that the optical paths are associated with position measurements at different instants, while the distances between the spacecraft $L_i$ are instantaneous.
By employing the obtained analytic expressions, the magnitudes of the residuals are further explored and discussed.

\begin{figure}[!t]
\includegraphics[width=0.40\textwidth]{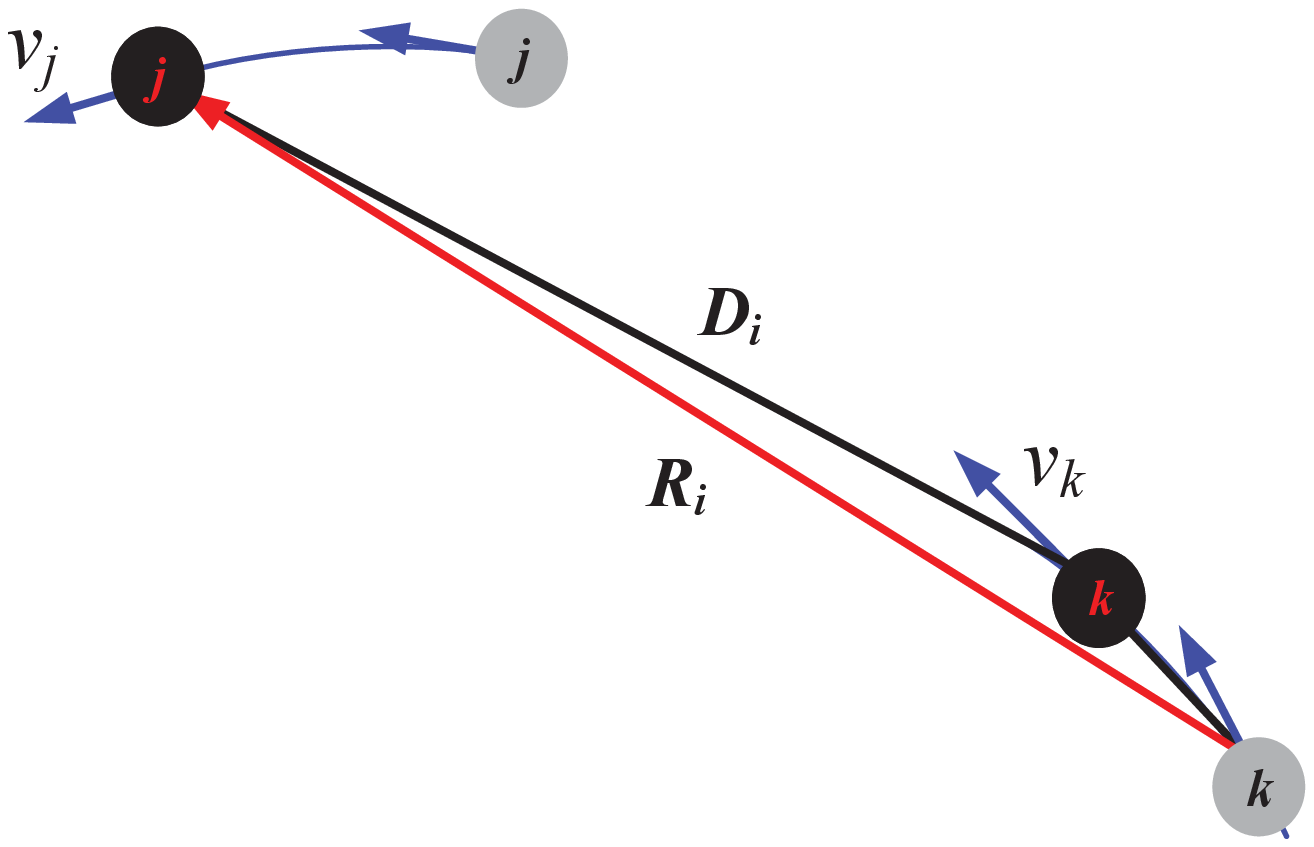}
\caption{\label{fig14}
The schematic diagram of the link $i$ defined by two spacecraft S/C $j$ and S/C $k$ in motion. 
The red vector $\boldsymbol{R}_i$ indicates that a laser beam was emitted by S/C $k$ at an earlier instant $t_1$ (indicated by gray filled circles) and received by S/C $j$ at a later instant $t_2$ (indicated by dark filled circles).
The black vector $\boldsymbol{D}_i$, on the other hand, represents the instantaneous vectorial difference between the two spacecraft at a later instant $t_2$.}
\end{figure}

As shown in Fig.~\ref{fig14}, the two spacecraft S/C $k$ and S/C $j$ are moving along their respective orbits.
The two gray-filled circles represent the positions of the two spacecraft at time $t_1$, and the black ones represent the positions at time $t_2$.
It is noted that the instantaneous distance $D_ i = |\boldsymbol{D}_ i| = |\boldsymbol{x}_j(t_2) - \boldsymbol{x}_k(t_2)|$ does not correspond the optical path of the laser beam.
The latter can be described by the coordinate distance $\boldsymbol{R}_ i= \boldsymbol{x}_j(t_2) - \boldsymbol{x}_k(t_1)$, and it can be expressed in terms of the former by Taylor expansion.
To be specific, one has~\cite{orbit-2014,orbit-2016}
\begin{align}\label{DISTAN}
\boldsymbol{R}_i = \boldsymbol{D}_i + \boldsymbol{v}_k (t_2)\left(\frac{{R}_i}{c}\right) - \frac{1}{2}{ \boldsymbol {a}_k}(t_2) \left(\frac{R_i}{c}\right)^2 + O({c^{ - 3}}) ,
\end{align}
where for $i$ (without prime), $j = i+1$ and $k = i-1$ are the S/C indices in the clock wise direction.  
The time interval $t_2-t_1 = {{R}_i}/{c} = {\left|\boldsymbol{R}_i\right|}/{c} \simeq \Delta t $, which has been substituted iteratively in order to obtain the expression.
Also, the corrections to the speed of light $c$ due to general relativity have been ignored in the adopted frame.

The module of the above quantity can be approximated by
\begin{align}\label{DISTAN11}
R_i = D_i + \frac{\boldsymbol{D}_i \cdot \boldsymbol{v}_k}{c}+ \frac{D_i}{2c^2}\left({v}_k^2 +\frac{(\boldsymbol{D}_i \cdot \boldsymbol{v}_k)^2}{ D_i^2} - \boldsymbol{D}_i\cdot \boldsymbol{a}_k \right) + O({c^{ - 3}}) ,
\end{align}
where the relevant quantites are evaluated at the moment that the light beam reaches the receiving spacecraft, $t = t_2$.

For an arbitrary link at $t=t_2$, regarding the first term of the r.h.s. of Eq.~\eqref{DISTAN11}, the optical path $L_i(t=t_2) \equiv R_i(t=t_2) \simeq D_i(t=t_2)$ can be expanded about the instant of measurement $t=t_0\equiv t_2 -\Delta T$, one has
\begin{align}\label{Diexp1}
D_i(t_2) &= D_i(t_0) + \hat{\boldsymbol{n}}_i\cdot (\boldsymbol{v_j}(t_0)-\boldsymbol{v_k}(t_0))(-\Delta T) \nonumber\\
&+ \frac{1}{2D_i}\left( (\boldsymbol{v_j}(t_0)-\boldsymbol{v_k}(t_0))\cdot (\boldsymbol{v_j}(t_0)-\boldsymbol{v_k}(t_0))+ D_i(t_0)\hat{\boldsymbol{n}}_i\cdot (\boldsymbol{a_j}(t_0)-\boldsymbol{a_k}(t_0))
-\left(\hat{\boldsymbol{n}}_i\cdot (\boldsymbol{v_k}(t_0)-\boldsymbol{v_j}(t_0))\right)^2  \right) (-\Delta T)^2  +  \cdots ,
\end{align}
where $\boldsymbol{D}_i = D_i \hat{\boldsymbol{n}}_i$.
It is further observed, when expanded about $t=t_0$, the second term of the r.h.s. of Eq.~\eqref{DISTAN11} gives the following contribution
\begin{align}\label{Diexp2}
\frac{\boldsymbol{D}_i(t_2) \cdot \boldsymbol{v}_k(t_2)}{c} = \frac{\boldsymbol{D}_i(t_0) \cdot \boldsymbol{v}_k(t_0)}{c} 
+ \left[\frac{(\boldsymbol{v}_j(t_0) - \boldsymbol{v}_k(t_0)) \cdot \boldsymbol{v}_k(t_0)}{c} +   \frac{\boldsymbol{D}_i(t_0) \cdot \boldsymbol{a}_k(t_0)}{c} \right](-\Delta T)+ \cdots.
\end{align}

Now, one can readily identify the specific forms of $L_i$ and $\dot{L}_i$ by substituting Eqs.~\eqref{Diexp1} and~\eqref{Diexp2} back into Eq.~\eqref{DISTAN11} and comparing against the Taylor expansion Eq.~\eqref{arm}.
To be specific, by extracting out the terms that are, respectively, constant and proportional to $\Delta T$, we have
\begin{align} \label{LNewDef}
L_i &= D_i + \frac{\boldsymbol{D}_i \cdot \boldsymbol{v}_k}{c} ,\nonumber\\
L_{i'} &= D_{i'} + \frac{\boldsymbol{D}_{i'} \cdot \boldsymbol{v}_j}{c} ,
\end{align}
and
\begin{align} \label{LdotNewDef}
\frac{\dot L_i}{c} &= \frac{\boldsymbol n_{i} \cdot (\boldsymbol v_j-\boldsymbol v_k)}{c} + \frac{1}{c^2}\left( (\boldsymbol v_j-\boldsymbol v_k) \cdot \boldsymbol v_k +\boldsymbol D_i \cdot \boldsymbol a_k \right) ,\nonumber\\
\frac{\dot L_{i'}}{c} &= \frac{\boldsymbol n_{i'} \cdot (\boldsymbol v_k-\boldsymbol v_j)}{c} + \frac{1}{c^2}\left( (\boldsymbol v_k-\boldsymbol v_j) \cdot \boldsymbol{v}_j +\boldsymbol{D}_{i'}\cdot \boldsymbol a_j \right) ,
\end{align}
where all the quantities on the r.h.s. of the expressions are evaluated at $t=t_0$.
It is noted that in geometric TDI, for the coefficients associated with $\dot{L}_i$ to be canceled out.
Subsequently, the difference between the two lines of Eq.~\eqref{LdotNewDef} reads
\begin{align}\label{differ2}
\frac{\dot L_i-\dot L_{i'}}{c} = \frac{1}{c^2}[\boldsymbol{v}_{j}^2- \boldsymbol{v}_{k}^2 + \boldsymbol{D}_{i}\cdot (\boldsymbol{a}_k + \boldsymbol{a}_j)],
\end{align}
where one utilizes $D_{j} = D_{i'}$ and $\boldsymbol{D}_{i} = \boldsymbol{x}_{j} - \boldsymbol{x}_{k}= - \boldsymbol{D}_{i'}$.

By making use of Eq.~\eqref{geoTDISecond}, as well as the values given in Tabs.~\ref{12linkTab},~\ref{14linkTab},~\ref{16linkTab},~\ref{16linkTab11}, and~\ref{16linkTab22}, we have $d_i = -d_{i'} \equiv d$.
Therefore, Eq.~\eqref{delaytt} can be simplified to read
\begin{align}\label{dimveo}
\sum\limits_{i = 1}^3 d_i{\frac{\dot L_i - \dot L_{i'}}{c}} =d\sum\limits_{i = 1}^3 {\frac{\dot L_i - \dot L_{i'}}{c}}  = \frac{d}{c^2}\sum\limits_{i = 1}^3\left[ \left( \boldsymbol{x}_{i+1} - \boldsymbol{x}_{i-1} \right) \cdot \left(\boldsymbol{a}_{i-1} + \boldsymbol{a}_{i+1} \right) \right] 
= \frac{d}{c^2}\left[ {\left(\boldsymbol{a}_j - \boldsymbol{a}_k \right)\cdot\left( \boldsymbol{x}_i - \boldsymbol{x}_k \right) + \left(\boldsymbol{a}_i - \boldsymbol{a}_k \right) \cdot \left( \boldsymbol{x}_k - \boldsymbol{x}_j \right)} \right] .
\end{align}
It is noted that the above quantity is merely a function of relative acceleration.

For the LISA detector, the position vectors of the three spacecraft under the first-order approximation are given by~\cite{orbit-2005}
\begin{align}\label{posi}
\boldsymbol{x}_k =e R\left[ [\cos{\Omega t - (k - 1)\frac{{2\pi }}{3}}],- 2\sin [ {\Omega t - (k - 1)\frac{{2\pi }}{3}}], \sqrt 3 \cos[ {\Omega t - (k - 1)\frac{{2\pi }}{3}}]\right] ,
\end{align}
where $\boldsymbol{x}_k=(x_k,y_k,z_k)$ represent the displacement between the individual S/C and the centers of the constellation and $e$ is the orbit eccentricity. 
The trajectory of the barycenter of three S/C is a circle in the ecliptic plane with a radius of $R$, armlength $L=2\sqrt{3}eR$, and the angular velocity $\Omega  = \sqrt {\frac{{GM}}{{R^3}}}$.
The acceleration of S/C $i$ is
 \begin{align}\label{accea}
 \boldsymbol a_i \approx -\frac{GM}{ {R_A}^3}\boldsymbol {R_A},
  \end{align}
where $R_A$ is the vector from S/C $i$ to the sun.
The acceleration of S/C $j$ can be written as
 \begin{align}\label{acceb}
 \boldsymbol a_j  \approx- \frac{GM}{\left| \boldsymbol {R_A} - \boldsymbol{D}_{k} \right|^3}\left( \boldsymbol {R_A} - \boldsymbol{D}_{k} \right).
 \end{align}
The acceleration of S/C $ k$ is written as
 \begin{align}\label{accec}
 \boldsymbol a_k  \approx -\frac{GM}{\left| \boldsymbol {R_A} + \boldsymbol{D}_{j} \right|^3}\left( \boldsymbol {R_A}+\boldsymbol{D}_{j} \right).
 \end{align}
According to Eq.~\eqref{dimveo}, only the relative acceleration is relevant, and it is assumed that $R_A\approx R$.

The Taylor expansions of the accelerations of S/C $j$ and S/C $k$ give
 \begin{subequations}
\begin{align}
- \boldsymbol a_j &\approx \frac{GM}{R^3}{\boldsymbol R} - \frac{GM}{R^3}{\boldsymbol{D}_k} - \frac{GM}{R^3}{\boldsymbol R}\frac{3\boldsymbol R \cdot (-\boldsymbol{D} _k)}{R^2} + 3\frac{GM}{R^3}{\boldsymbol{D} _k}\frac{D_k\cos \theta _k(t)}{R} + \frac{3}{2}\frac{GM}{R^3}{\boldsymbol R}\frac{D _{k}^2}{R^2}\left( 5\cos^2 \theta _k(t) - 1 \right),\label{acce11}\\
- \boldsymbol a_k &\approx \frac{GM}{R^3}{\boldsymbol R} + \frac{GM}{R^3}{\boldsymbol{D}_j} - \frac{GM}{R^3}{\boldsymbol R}\frac{3\boldsymbol R \cdot \boldsymbol{D}_j}{R^2} - 3\frac{GM}{R^3}{\boldsymbol{D}_j}\frac{D _j\cos \theta _j(t)}{R} + \frac{3}{2}\frac{GM}{R^3}{\boldsymbol R}\frac{D_{j}^2}{R^2}\left( 5\cos^2 \theta _j(t) - 1 \right).\label{acce12}
\end{align}
\end{subequations}
Here, we have denoted the angle between $\boldsymbol R$ and $-\boldsymbol{D}_{k}$ by ${\theta _k}(t)$, 
and the angle between $\boldsymbol R$ and $\boldsymbol{D}_j$ by ${\theta _j}(t)$.

By combining Eqs.~\eqref{dimveo},~\eqref{acce11} and~\eqref{acce12}, one finds
\begin{align}\label{vecresult}
\sum\limits_{i{\rm{ = 1}}}^3 d_i {\frac{\dot{L}_i - \dot{L}_{i'}}{c}}  = \frac{d}{c^2}
\frac{3}{2}\frac{GM}{R^3}\frac{L^3}{R}[ \cos {\theta _j}(t) - \cos {\theta _k}(t)]
 + \frac{d}{c^2}\frac{3}{2}\frac{GM}{R^3}\frac{L^3}{R}[( 5\cos ^2{\theta _k}(t) - 1 )\cos {\theta _j}(t) -( 5\cos ^2{\theta _j}(t) - 1)\cos {\theta _k}(t)] ,
\end{align}
which can be further simplified to
\begin{align}\label{vecresultsim}
\sum\limits_{i{\rm{ = 1}}}^3 d_i{\frac{\dot{L}_i - \dot{L}_{i'}}{c}}  = \frac{d}{c^2}\frac{{15}}{2}\frac{{GM}}{{R^3}}\frac{{{L^3}}}{{{R}}}\cos {\theta _j}(t)\cos {\theta _k}(t)\left[ {\cos {\theta _k}(t) - \cos {\theta _j}(t)} \right].
\end{align}
 
To evaluate $\theta _i,\theta _k$, we need to project $\boldsymbol{D}_{j},\boldsymbol{D}_{k}$ to the $x$-axis, subsequently we have
\begin{align}\label{theta}
\cos {\theta _j} =&  - \frac{1}{2}\sin \left( {\Omega t + \frac{\pi }{3}} \right),\\\notag
\cos {\theta _k} =& \frac{1}{2}\sin \left( {\Omega t - \frac{\pi }{3}} \right),
\end{align}
where $\Omega$ is the average angular velocity.
By substituting Eq.~\eqref{theta} into Eq.~\eqref{vecresultsim}, the optical path discrepancy associated with $\dot{L}_i$ term is found to be
\begin{align}\label{deltvec}
\delta {t_v} \approx d \frac{L}{c}\frac{{15}}{{64}}\frac{GML^3}{R^4c^2}\sin 3\Omega t.
\end{align}

Similarly, one can evaluate Eq.~\eqref{delaytt2}.
For the optimized case, the expression of $\ddot{L}_i$ is~\cite{orbit-2005,orbit-2019}:
\begin{align}\label{ACCE} 
{\ddot {L}_i} = R{\Omega ^2}\left( {\frac{{15\sqrt 3 }}{{16}}\cos {\theta _i} + \frac{{9\sqrt 3 }}{{16}}\cos 3{\theta _i}} \right){e^2},
\end{align}
where ${\theta _i} = \Omega t - \frac{\pi }{3} - i\frac{{2\pi }}{3}$.
Utilizing the above result one finds
\begin{align}\label{dimeacce} 
\sum\limits_{i = 1}^3 {{f_i}} \frac{\ddot {L}_i L}{c^2} = R{\Omega ^2}\sum\limits_{i = 1}^3 {{f_i}} \left( {\frac{{15\sqrt 3 }}{{16}}\cos {\theta _i} + \frac{{9\sqrt 3 }}{{16}}\cos 3{\theta _i}} \right){e^2}\frac{L}{{{c^2}}} ,
\end{align}
where $f_i$ are the coefficients of the acceleration terms in Eqs.~\eqref{delat}.
In particular, it is observed that the coefficients $\{f_1, f_2, f_3\}$ satisfy $f_1+f_2+f_3=0$.
Therefore, the coefficient before the term $\cos 3{\theta _i}$ in Eq.~\eqref{dimeacce} always vanishes, and the resulting optical path discrepancy associated with $\ddot{L}_i$ term is given by
\begin{align}\label{deltaacce}
\delta t_a \approx \frac{L}{c}\frac{5\sqrt{3}}{128}\frac{GML^3}{R^4c^2}\sum\limits_{i = 1}^3 {{f_i}} \cos {\theta _i}.
\end{align}

By summing up Eqs.~\eqref{deltvec} and~\eqref{deltaacce}, the delay-time residual reads
\begin{align}\label{deltatt}
\delta t \approx d \frac{L}{c}\frac{{15}}{{64}}\frac{GML^3}{R^4c^2}\sin 3\Omega t +\frac{L}{c}\frac{5\sqrt{3}}{128}\frac{GML^3}{R^4c^2}\sum\limits_{i = 1}^3 {{f_i}} \cos {\theta _i} .
\end{align}
The coefficients $d, f_i$ and the resultant expressions for the residual can be found in Tabs.~\ref{12linkTab},~\ref{14linkTab},~\ref{16linkTab},~\ref{16linkTab11}, and~\ref{16linkTab22}.
As a rough estimation, one employs the following typical values of the parameters.
The instantaneous distance between the spacecraft is $D_{k} \approx D_{j} \approx D_{i} \equiv L \approx 2.5\times 10^9{\rm m}$,
the acceleration of the sun is $6\times10^{-3}{\rm m/s^2}$,
and the distance between the center of the constellation and the sun is $R \approx 1.5\times 10^{11}{\rm m}$.
Then, according to Eq.~\eqref{deltatt}, for the modified second-generation TDI combinations, the residual only receives a contribution from the second term of Eq.~\eqref{deltatt}, when compared to their second-generation counterparts. 
Using the above specific parameters, the coefficient of the term $\sin 3\Omega t$ is approximately $d\times 9\times10^{-14}\rm{s}$,
and the coefficient of the term $\cos \Omega t$ is $\sum\limits_{i = 1}^3 {f_i}\times 4.5\times10^{-14}\rm{s}$.
In the last two columns of the tables, we estimate the magnitude for different frequency doubling terms (where the initial phase has been neglected) in the delay-time residuals.
It is observed that the mismatch of the optical paths for both second-generation and modified second-generation TDI combinations is a few pico-seconds. 
The magnitude of the obtained results is consistent with those from numerical simulations~\cite{tdi-geometric-2020}.

\section{Sensitivity function}\label{section5}

In this section, we explore the response functions, residual noise PSDs, and sensitivity curves of the obtained geometric TDI combinations by making use of the analytical results derived in~\cite{response-full-03}.  
These formulae provide straightforward access to the analytic characteristics of the spaceborne GW detector.

The averaged response function for an arbitrary TDI combination reads
\begin{align}\label{analytical}
R(u) &= \frac{2}{4}C_1[\tilde P_i(u)]\times{f_1}(u) +C_2[\tilde P_i(u)]\times{f_2}(u)+\frac{3}{4}C_3[\tilde P_i(u)]\times{f_3}(u)-\frac{3}{4}C_4[\tilde P_i(u)]\times{f_4}(u)+\frac{1}{4}C_5[\tilde P_i(u)]\times{f_5}(u) ,
\end{align}
where the coefficients $\tilde P_i(u)$ are the Fourier transforms of $P_i$, polynomials of the delay operators, given in Eq.~\eqref{tdi}. 
It is noted that the Fourier transform of the delay operator possesses the form ${\tilde{\cal D}}_{i}=e^{iu}$, where one defines the dimensionless quantity ${u=\frac{2\pi f L}{c}}$.
The coefficient $C_i[\tilde P_i(u)]$ and $f_i(u)$ are given by
\begin{align}\label{coff}
C_1[\tilde P_i(u)]&=\sum\limits_{i = 1}^3[\tilde P_i|^2 + |{\tilde P_{i^\prime}}|^2],\notag\\
C_2[\tilde P_i(u)]&=2\sum\limits_{i = 1}^3{\rm Re}[{\tilde P_i}{\tilde P_{(i + 1)^\prime}}^{*}],\notag\\
C_3[\tilde P_i(u)]&=2\sum\limits_{i = 1}^3{\rm Re}[({\tilde P_i}{\tilde P_{i + 1}}^{\rm{*}} + {\tilde P_{i^\prime}} {\tilde P_{(i - 1)^\prime}}^{*}){e^{iu}}],\notag\\
C_4[\tilde P_i(u)]&= 2\sum\limits_{i = 1}^3{\rm Im}[({\tilde P_i}{\tilde P_{i + 1}}^{\rm{*}} + {\tilde P_{i^\prime}} {\tilde P_{(i - 1)^\prime}}^{*}){e^{iu}}],\notag\\
C_5[\tilde P_i(u)]&=2\sum\limits_{i = 1}^3{\rm Re}[{\tilde P_i}{\tilde P_{i^{\prime}}}^{*} + {\tilde P_i}{\tilde P_{(i - 1)^\prime}}^{*}],
\end{align} 
and
\begin{align}\label{fu}
{f_1}(u) &= \frac{4}{3} - \frac{2}{{{u^2}}} + \frac{{\sin 2u}}{{{u^3}}},\notag\\
{f_{\rm{2}}}(u)& = \frac{{ - u\cos u + \sin u}}{{{u^3}}} - \frac{{\cos u}}{3},\notag\\
{f_{\rm{3}}}(u) &\!=\! \log \frac{4}{3}\! \!-\!\! \frac{{\rm{5}}}{{{\rm{18}}}}\!\!+\!\!\frac{{ \!-\! {\rm{5}}\sin u{\rm{ \!+\! 8}}\sin 2u \!-\! 3\sin 3u}}{{8u}} \!-\! \frac{{\rm{1}}}{{\rm{3}}}\left(\! {\frac{{{\rm{4 \!+\! 9}}\cos u\! + \!{\rm{12}}\cos 2u{\rm{ \!+\! }}\cos 3u}}{{{\rm{8}}{u^{\rm{2}}}}}}\! \right)
\!\!+\!\!\frac{{\rm{1}}}{{\rm{3}}}\left(\! {\frac{{ \!-\! {\rm{5}}\sin u{\rm{ \!+\! 8}}\sin 2u{\rm{ \!+\! 5}}\sin {\rm{3}}u}}{{{\rm{8}}{u^3}}}}\! \right)\! \!+\!\! {\rm{Ci3}}u \!\!-\!\! {\rm{2Ci}}2u\! \!+\!\! {\rm{Ci}}u,\notag\\
{f_{\rm{4}}}(u) &= \frac{{ - {\rm{5}}\cos u{\rm{ + 8}}\cos 2u - {\rm{3}}\cos 3u}}{{{\rm{8}}u}} + \frac{{\rm{1}}}{{\rm{3}}}\left( {\frac{{{\rm{9}}\sin u{\rm{ + 12}}\sin {\rm{2}}u + \sin 3u}}{{{\rm{8}}{u^2}}} - \frac{{{\rm{8 + 5}}\cos u - {\rm{8}}\cos 2u - {\rm{5}}\cos 3u}}{{{\rm{8}}{u^3}}}} \right)
{\rm{ + }}2{\rm{Si}}2u - {\rm{Si3}}u - {\rm{Si}}u,\notag\\
{f_{\rm{5}}}(u) &=  - \log 4{\rm{ + }}\frac{7}{6} + \frac{{{\rm{11}}\sin u - {\rm{4sin2}}u}}{{4u}} - \frac{{{\rm{10 + 5}}\cos u - {\rm{2cos2}}u}}{{{\rm{4}}{u^{\rm{2}}}}}{\rm{ + }}\frac{{{\rm{5}}\sin u + {\rm{4}}\sin 2u}}{{{\rm{4}}{u^{\rm{3}}}}} + {\rm{2}}\left( {{\rm{Ci2}}u - {\rm{Ci}}u} \right){\rm{.}}
\end{align}
Here, one denotes the functions {\it SinIntegral} by ${\rm{Si}}(z) = \int_0^z {\sin t/tdt}$ and {\it CosIntegral} by ${\rm{Ci}}(z) =  - \int_z^\infty  {\cos t/tdt}$.

The total noise PSD can be written as
\begin{align}\label{noise}
N(u)&=S_{\rm{TDI}^a}(u )+S_{{\rm{TDI}}^{\rm{shot}}}(u)\notag\\
    &=C_1[\tilde P_i(u)]n_1(u)+ 2C_2[\tilde P_i(u)]n_2(u) ,
\end{align}
and
\begin{align}\label{nu}
n_1(u) &= 2 \frac{L^2 s_a^2 }{u^2 c^4}+\frac{u^2 s_{x}^2}{L^2},\notag\\
n_2(u)& = \frac{ L^2 s_a^2}{u^2 c^4}\cos u .
\end{align}

In what follows, the calculations are carried out using the LISA mission's typical parameters.
The corresponding amplitude spectral densities of the test mass and shot noise are, respectively, ${s_{a}^{\rm{LISA}}=3\times10^{-15}\rm{ms^{-2}/\sqrt{\rm{Hz}}}}$ and ${s_{x}^{\rm{LISA}}=10\times10^{-12}\rm{m/\sqrt{\rm{Hz}}}}$.

For the sixteen-link combinations, we first analyze the modified second-generation TDI combinations.
Among the nine solutions, four of them are the standard forms $[X]_1^{16}, [U]_1^{16}, [E]_1^{16}, [P]_1^{16}$.

For the Michelson-type TDI combinations, by substituting Eq.~\eqref{x1coff} and Eq.~\eqref{x2coff} into Eqs.~\eqref{analytical}-\eqref{nu}, the GW signal response function and the noise PSD of $[X]_1^{16}$ are given by
\begin{align}\label{x1response}
R(u)_{[X]_1^{16}} =& \frac{8}{3}{\sin ^2}u{\sin ^2}2u\{
5 + \cos 2u + 12\left( {{\rm{Ci}}u - {\rm{Ci}}2u + \log 2} \right)-18\cos 2u( {{\rm{Ci}}u - 2{\rm{Ci}}2u + {\rm{Ci}}3u + \log \frac{4}{3}})\\\notag
 -& 18\sin 2u\left( {{\rm{Si}}u - 2{\rm{Si2}}u + {\rm{Si3}}u} \right) + \frac{{3\left( { - 7\sin u + 2\sin 2u} \right)}}{u} - \frac{{3\cos u\left( { - 5 + 8\cos u} \right)}}{{{u^2}}} + \frac{{3\left( { - 5\sin u + 4\sin 2u} \right)}}{{{u^3}}}\},
\end{align}
and
\begin{align}\label{x1noise}
N(u)_{[X]_1^{16}} = 64{\sin^2}2u{\sin ^2}u\left[ {\left( {3 + \cos 2u} \right)\frac{{{L^2}{s_a}^2}}{{{u^2}{c^4}}} + \frac{{{u^2}{s_x}^2}}{{{L^2}}}} \right].
\end{align}
Also, the GW signal response function and the noise PSD of the alternative form $[X]_2^{16}$ are found to be
\begin{align}\label{x2response}
R(u)_{[X]_2^{16}} = \frac{R(u)_{[X]_1^{16}}}{\sin^2 2u}\sin^2 u,
\end{align}
and
\begin{align}\label{x2noise}
N(u)_{[X]_2^{16}} = \frac{N(u)_{[X]_1^{16}}}{\sin^2 2u}\sin^2 u.
\end{align}

\begin{figure}[!t]
\begin{tabular}{cc}
\vspace{0pt}
\begin{minipage}{225pt}
\centerline{\includegraphics[width=200pt]{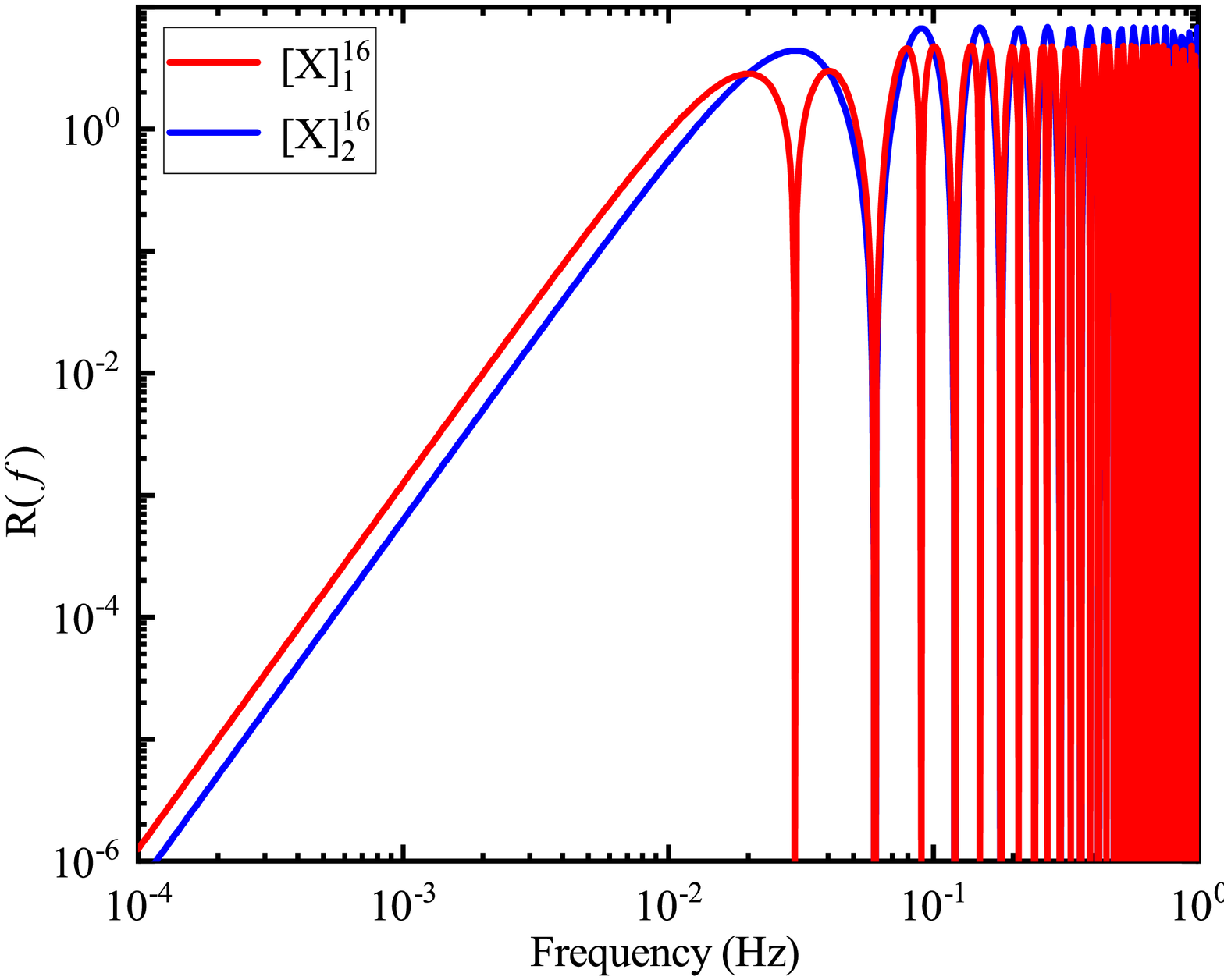}}
\end{minipage}
&
\begin{minipage}{225pt}
\centerline{\includegraphics[width=200pt]{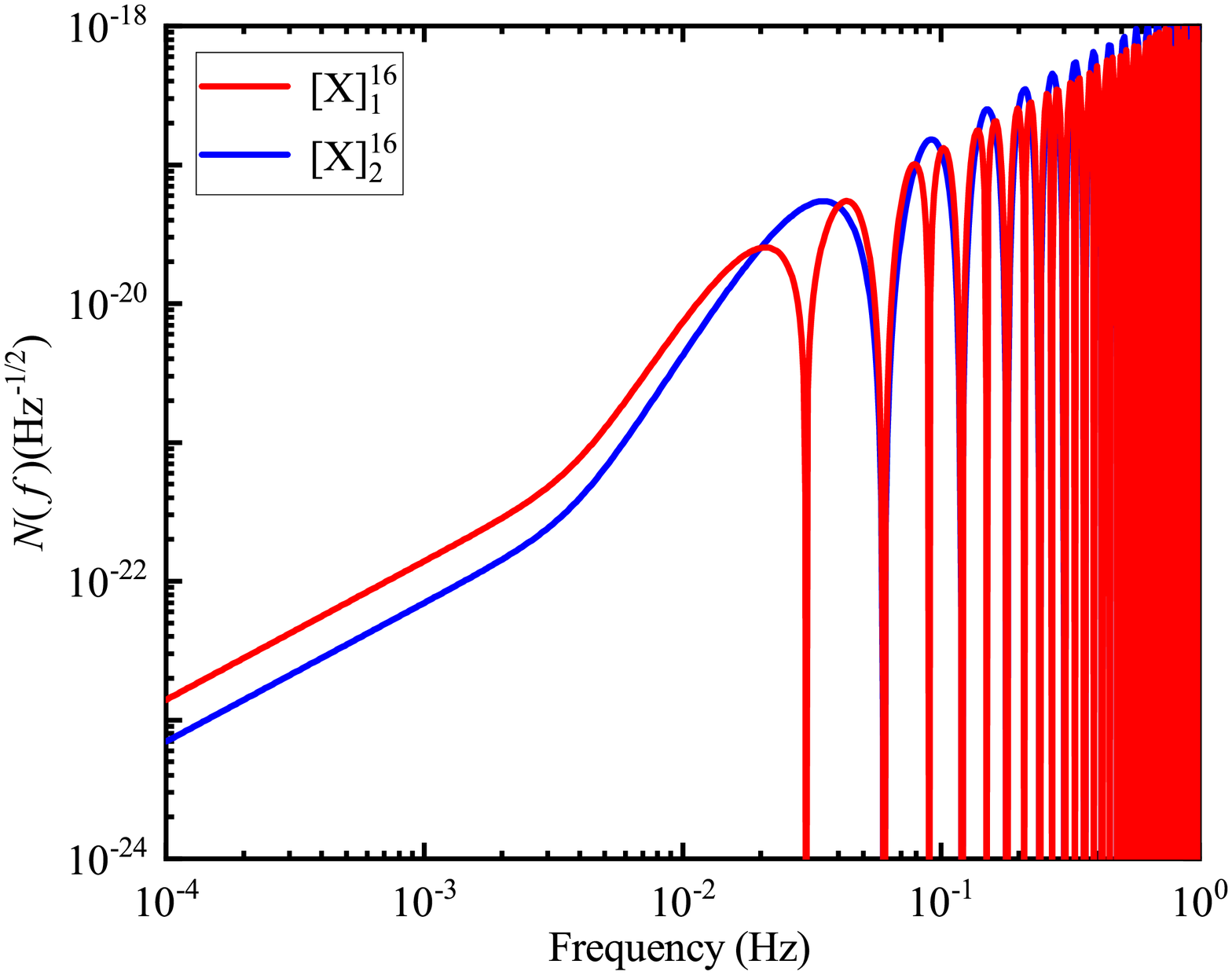}}
\end{minipage}
\end{tabular}
\renewcommand{\figurename}{Fig.}
\caption{\label{fig15}
The GW averaged response function and noise PSD of the sixteen-link Michelson-type TDI combinations.
The calculations were carried out using the parameters of the LISA detector.
The left panel shows the  GW averaged response function, while the right panel shows
that for the noise PSD.}
\end{figure}

The resulting GW response function and noise PSD for the two Michelson-type combinations $[X]_1^{16}$ and $[X]_2^{16}$ are shown in Fig.~\ref{fig15}.
Here, the zeros of the functions can be identified using Eqs.~\eqref{x1response}-\eqref{x2noise}.
To be specific, the zeros of the alternative form $[X]_2^{16}$, governed by $\sin u=0$. 
They are more sparsely distributed, located at $nc/2L$, where $n$ is an integer.
On the other hand, the zeros of the standard form $[X]_1^{16}$ are at $nc/4L$, determined by $\sin 2u=0$.
These results are consistent with the observations given by Vallisneri~\cite{tdi-geometric-2005}.

For the Relay-type TDI combinations, the response functions and noise PSD are obtained by substituting Eq.~\eqref{u1coff}, Eq.~\eqref{u2coff} and Eq.~\eqref{u3coff} into Eqs.~\eqref{analytical}-\eqref{nu}.
For the standard form $[U]_1^{16}$, we have
\begin{align}\label{u1response}
R(u)_{[U]_1^{16}}&= \frac{1}{3}(1 + 2\cos u)^2{\sin ^4}\frac{u}{2}\{72 + 56\cos u + 16\cos 2u- \frac{3(108\sin u - 8\sin 2u - 8\sin 3u + \sin 4u)}{u}\\\notag
&- \frac{3(29 - 12\cos u + 34\cos 2u + 16\cos 3u + 5\cos 4u)}{u^2}+ \frac{3(- 44\sin u + 24\sin 2u + 16\sin 3u + 5\sin 4u)}{u^3}\\\notag
&+ 48[(7 + 6\cos u - \cos 2u){\rm{Ci}}u - 2(5 + 3\cos u - 2\cos 2u){\rm{Ci}}2u 
+\cos 2u\log \frac{27}{16} + \log \frac{1024}{27} + \cos u\log 64 + 6{\rm{Ci}}3u{\sin^2}u \\\notag
&-3(\sin u + \sin 2u)({\rm{Si}}u - 2{\rm{Si2}}u + {\rm{Si3}}u)]\},
\end{align}
and
\begin{align}\label{u1noise}
N(u)_{[U]_1^{16}} = 32{\left( {1 + 2\cos u} \right)^2}{\sin ^4}\frac{u}{2}[2\left( {5 + 5\cos u + 2\cos 2u} \right) \times \frac{{{L^2}{s_a}^2}}{{{u^2}{c^4}}}
 + \left( {4 + 4\cos u + \cos 2u} \right) \times \frac{u^2 s_x^2}{L^2}].
\end{align}

For the alternative form $[U]_2^{16}$, one finds
\begin{align}\label{u2response}
 R(u)_{[U]_2^{16}}=\frac{R(u)_{[U]_1^{16}}}{(1 + 2\cos u)^2\sin^2 \frac{u}{2}}{\sin ^2}u,
\end{align}
and
\begin{align}\label{u2noise}
 N(u)_{[U]_2^{16}}=\frac{N(u)_{[U]_1^{16}}}{(1 + 2\cos u)^2\sin^2 \frac{u}{2}}{\sin ^2}u.
\end{align}
 Also, the expression for the GW signal response function and the noise PSD of the alternative form $[U]_3^{16}$ are found to be
\begin{align}\label{u3response}
 R(u)_{[U]_3^{16}}=\frac{R(u)_{[U]_1^{16}}}{(1 + 2\cos u)^2},
\end{align}
and
\begin{align}\label{u3noise}
 N(u)_{[U]_3^{16}}=\frac{N(u)_{[U]_1^{16}}}{(1 + 2\cos u)^2}.
\end{align}

\begin{figure}[!t]
\begin{tabular}{cc}
\vspace{0pt}
\begin{minipage}{225pt}
\centerline{\includegraphics[width=200pt]{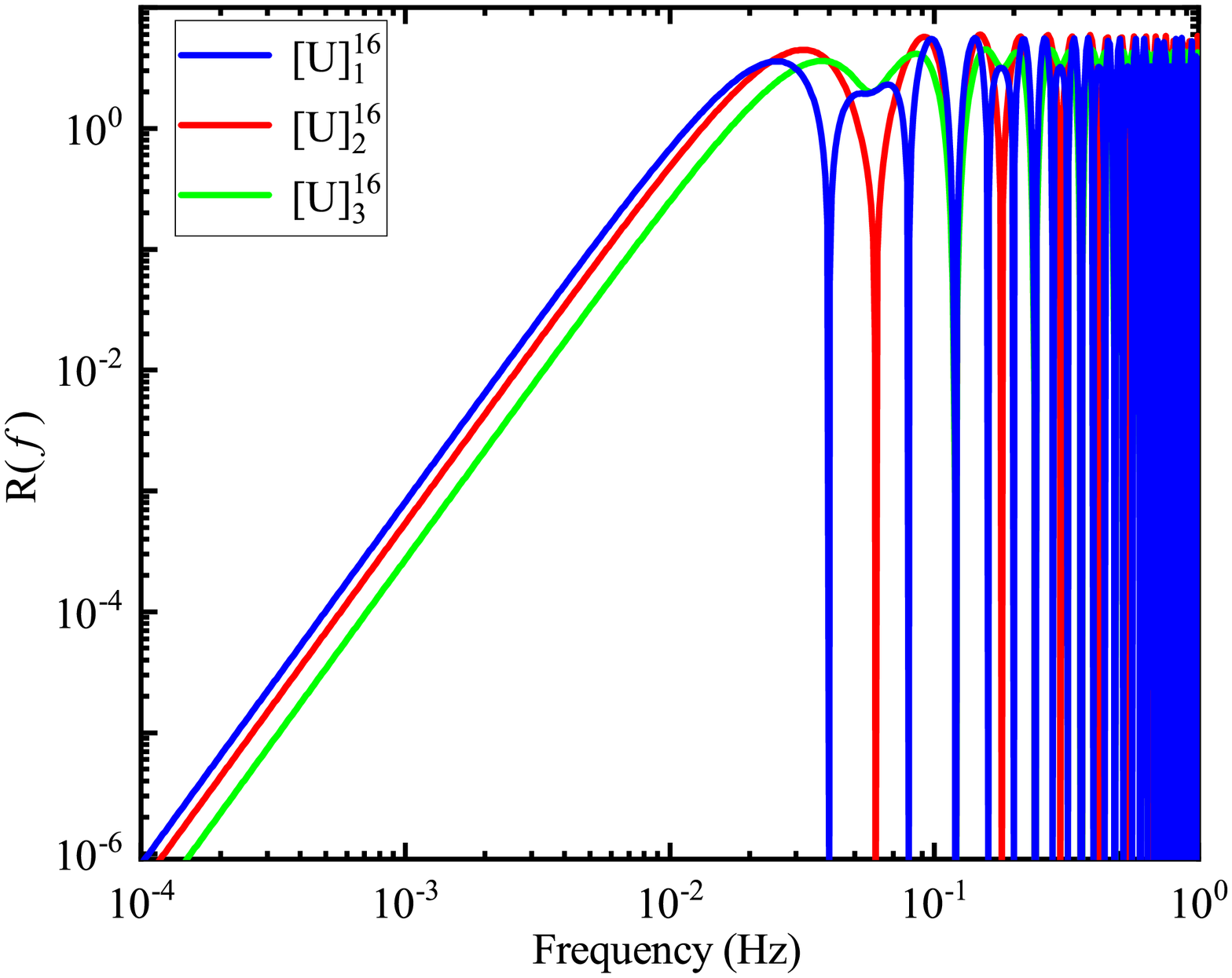}}
\end{minipage}
&
\begin{minipage}{225pt}
\centerline{\includegraphics[width=200pt]{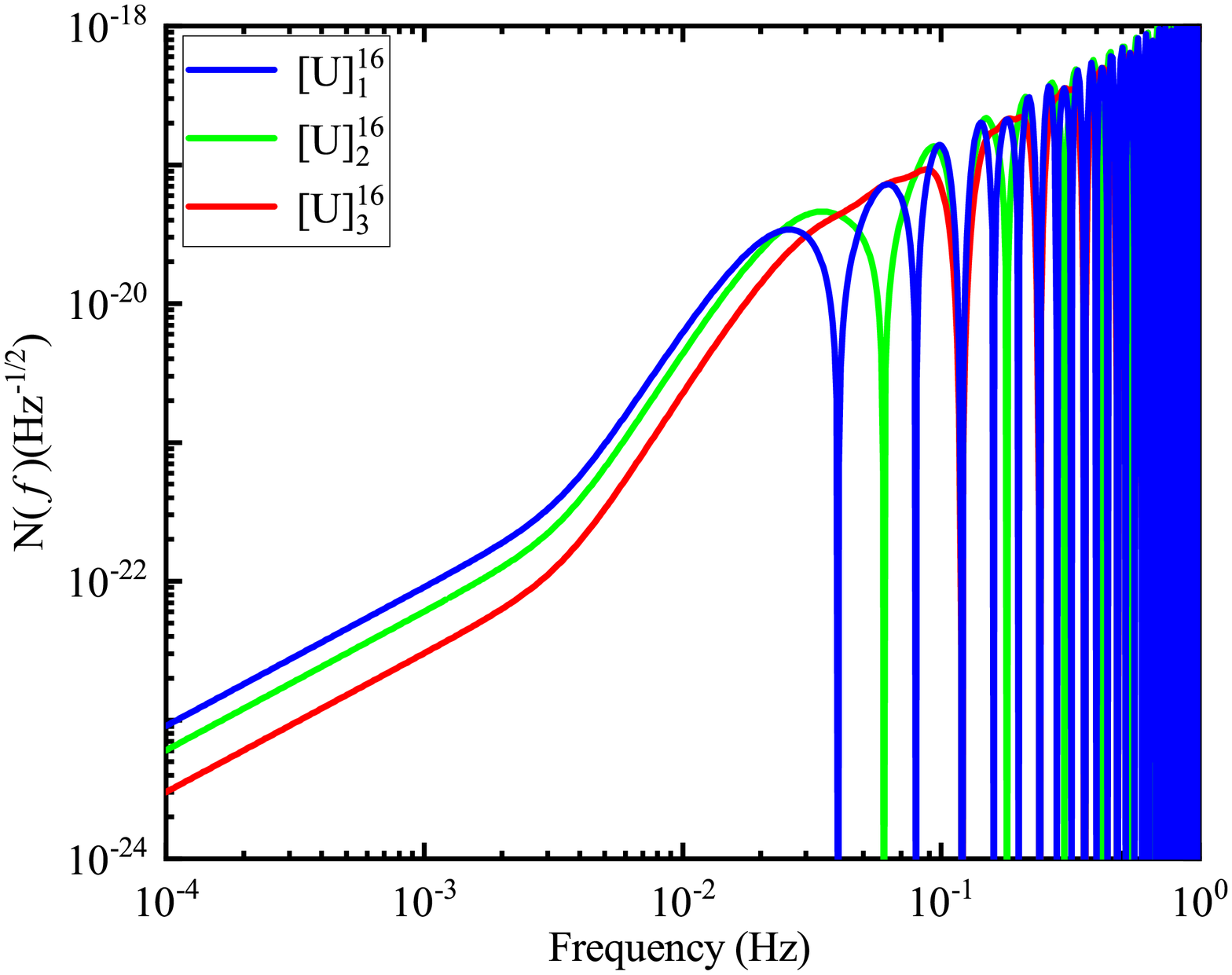}}
\end{minipage}
\end{tabular}
\renewcommand{\figurename}{Fig.}
\caption{\label{fig16}
The GW averaged response function and noise PSD of the sixteen-link Relay-type TDI combinations.
The calculations were carried out using the parameters of the LISA detector.
The left panel shows the  GW averaged response function, while the right panel shows that for the noise PSD.}
\end{figure}

Again, the zeros of the GW averaged response function, and noise PSD can be analyzed using the analytic forms given by Eqs.~\eqref{u1response}-\eqref{u3noise}.
To be specific, for standard form $[U]_1^{16}$, they are located at $nc/3L$, governed by $\cos u=-\frac{1}{2}$.
On the other hand, those of the alternative forms $[U]_2^{16}$ and $[U]_3^{16}$ are at $nc/2L$ and $nc/L$, respectively, determined by $\sin u=0$ and $\sin \frac{u}{2}=0$.
The above functions are shown in Fig.~\ref{fig16}.

For the Monitor-type TDI combinations, by substituting Eq.~\eqref{e1coff} and Eq.~\eqref{e2coff} into Eqs.~\eqref{analytical}-\eqref{nu}, the GW signal response function and the noise PSD of the standard form $[E]_1^{16}$ are given by
\begin{align}\label{e1response}
R(u)_{[E]_1^{16}} =& \frac{2}{3}{\sin ^2}u{\sin ^2}\frac{u}{2}\{
(20 + 4\cos u + 168\left( {1 + \cos u} \right){\rm{Ci}}u - 240\left( {1 + \cos u} \right){\rm{Ci}}2u\\\notag
 -& \frac{{3\left( {27\sin u - 9\sin 2u + \sin 3u} \right)}}{u} - \frac{{3\left( {11 - 15\cos u + 11\cos 2u + 5\cos 3u} \right)}}{{{u^2}}}\\\notag
 + &\frac{{3\left( { - 25\sin \left[ u \right] + 11\sin \left[ {2u} \right] + 5\sin \left[ {3u} \right]} \right)}}{{{u^3}}}\\\notag
 +& 24[ {3\left( {1 + \cos u} \right){\rm{Ci}}3u + \left( {1 + \cos u} \right)\log \frac{{1024}}{{27}} + 3\sin u\left( {{\rm{Si}}u - 2{\rm{Si2}}u + {\rm{Si3}}u} \right)}]\},
\end{align}
and
\begin{align}\label{e1noise}
N(u)_{[E]_1^{16}} = 32{\sin ^2}u{\sin ^2}\frac{u}{2}\left[ {2\left( {3 + \cos u} \right) \times \frac{{{L^2}{s_a}^2}}{{{u^2}{c^4}}} + \left( {3 + 2\cos u} \right) \times \frac{{{u^2}{s_x}^2}}{{{L^2}}}} \right].
\end{align}

Those for the alternative form $[E]_2^{16}$ are found to be
\begin{align}\label{e2response}
R(u)_{[E]_2^{16}}=\frac{R(u)_{[E]_1^{16}}}{\sin^2 u}{\sin^2\frac{u}{2}},
\end{align}
and
\begin{align}\label{e2noise}
 N(u)_{[E]_2^{16}}=\frac{N(u)_{[E]_1^{16}}}{\sin^2 u}{\sin^2\frac{u}{2}}.
\end{align}

For the Beacon-type TDI combinations, by substituting Eq.~\eqref{p1coff} and Eq.~\eqref{p2coff} into Eqs.~\eqref{analytical}-\eqref{nu}, the GW signal response function and noise PSD of the standard form $[P]_1^{16}$ are given by
\begin{align}\label{p1resnoise}
R(u)_{[P]_1^{16}} =R(u)_{[E]_1^{16}},N(u)_{[P]_1^{16}} =N(u)_{[E]_1^{16}},
\end{align}
and those for the alternative form $[P]_2^{16}$ are found to be
\begin{align}\label{p2resnoise}
R(u)_{[P]_2^{16}} =R(u)_{[E]_2^{16}},N(u)_{[P]_2^{16}} =N(u)_{[E]_2^{16}}.
\end{align}

\begin{figure}[!t]
\begin{tabular}{cc}
\vspace{0pt}
\begin{minipage}{225pt}
\centerline{\includegraphics[width=200pt]{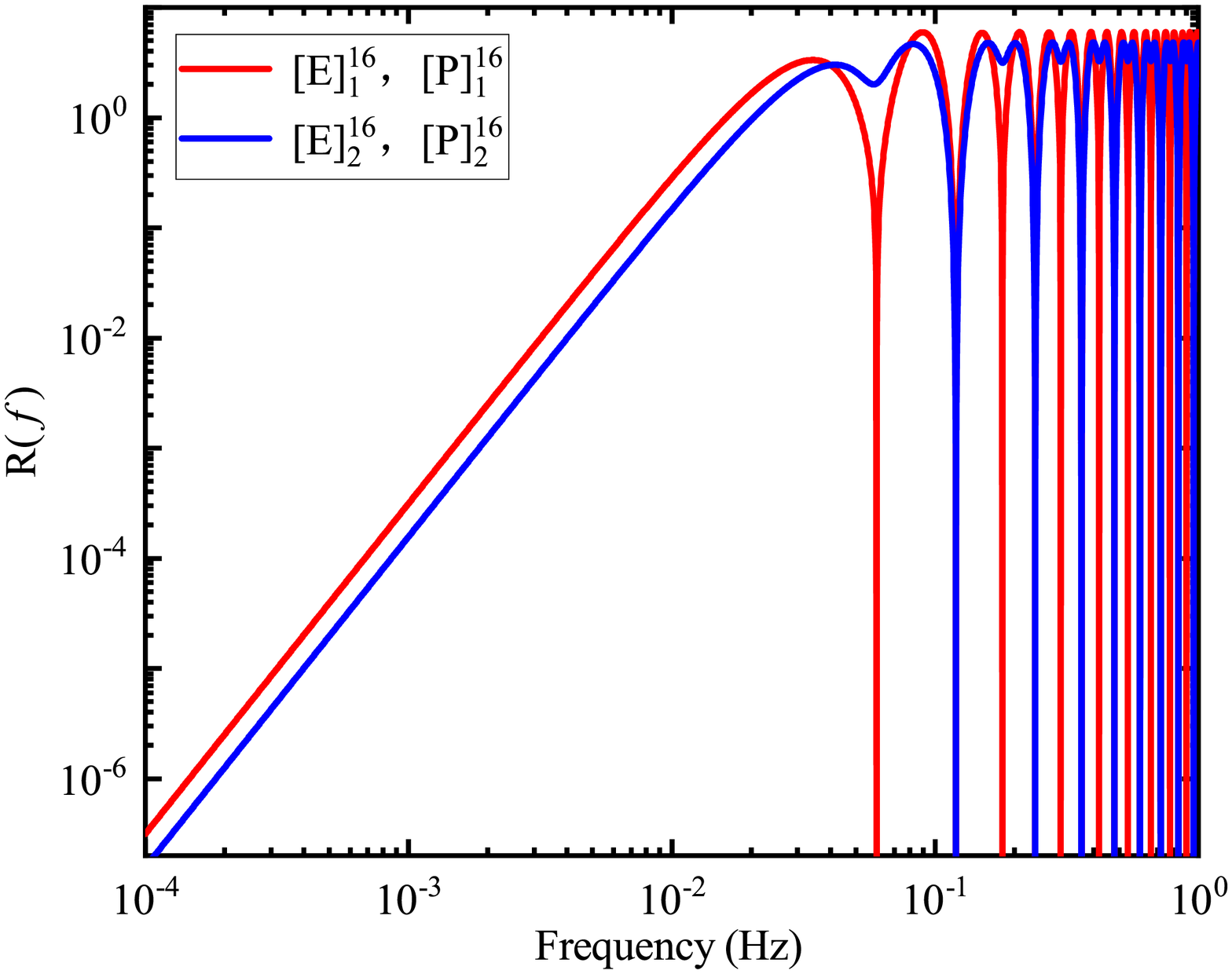}}
\end{minipage}
&
\begin{minipage}{225pt}
\centerline{\includegraphics[width=200pt]{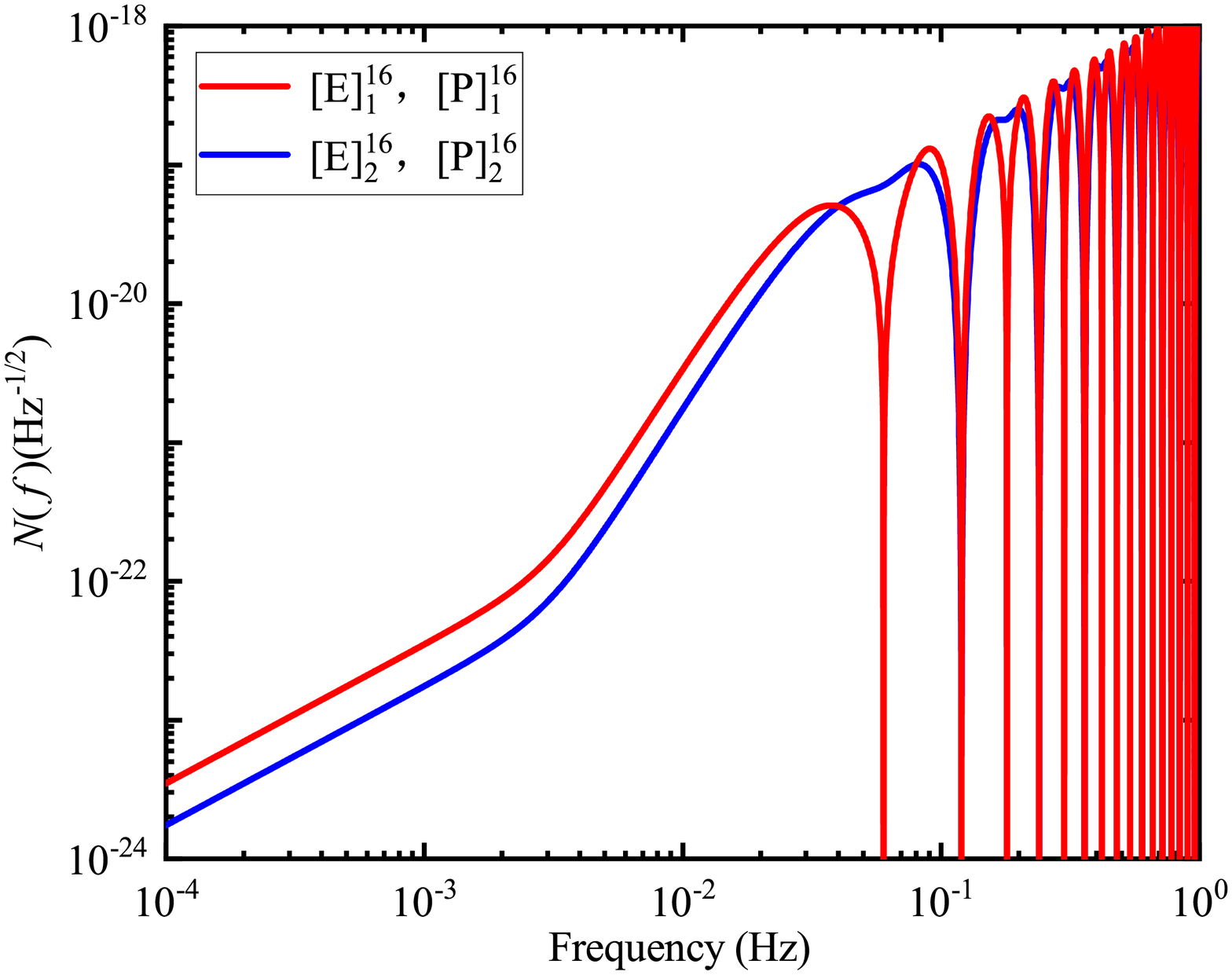}}
\end{minipage}
\end{tabular}
\renewcommand{\figurename}{Fig.}
\caption{\label{fig17}
The GW averaged response function and noise PSD of the sixteen-link  Monitor-type and Beacon-type TDI combinations.
The calculations were carried out using the parameters of the LISA detector.
The left panel shows the  GW averaged response function, while the right panel shows
that for the noise PSD.}
\end{figure}

The zeros of the GW averaged response function, and noise PSD can be derived using Eqs.~\eqref{e1response}-\eqref{p2resnoise}.
The zeros of the standard form $[E]_1^{16}$ and $[P]_1^{16}$ are located at $nc/2L$, in accordance with $\sin u=0$.
Those of the alternative forms $[E]_2^{16}$ and $[P]_2^{16}$ are at $nc/L$, governed by $\sin \frac{u}{2}=0$.
The resulting functions are presented in Fig.~\ref{fig17}.

In terms of the response function and noise PSD, the sensitivity function is defined as
\begin{align}\label{senfun}
S(u) = \sqrt{\frac{N(u)}{\frac{2}{5}R(u)}} ,
\end{align}
where the factor $2/5$ is the average of orbit inclination for the binary system.
It is interesting to note that the alternative form of the Michelson-type combination $[X]_2^{16}$ possesses the same sensitivity function as the standard form $[X]_1^{16}$
\begin{align}\label{x1x2senfun}
S(u)_{[X]_1^{16}} = S(u)_{[X]_2^{16}}.
\end{align}
However, the GW response function and noise PSD of $[X]_2^{16}$ have fewer zeros.

The alternative forms of the Relay-type combinations, $[U]_2^{16}$ and $[U]_3^{16}$, also have an identical sensitivity function as the standard form $[U]_1^{16}$
\begin{align}\label{U1U2U1U3senfun}
S(u)_{[U]_1^{16}} = S(u)_{[U]_2^{16}} =S(u)_{[U]_3^{16}}.
\end{align}

Similarly, for the Monitor-type and Beacon-type TDI combinations, it is found that the alternative TDI combinations possess the same sensitivity function as the standard ones
\begin{align}\label{EPsenfun}
S(u)_{[E]_1^{16}} =S(u)_{[P]_1^{16}}=S(u)_{[E]_2^{16}} =S(u)_{[P]_2^{16}}.
\end{align}

\begin{figure}[!t]
\begin{tabular}{cc}
\vspace{0pt}
\begin{minipage}{225pt}
\centerline{\includegraphics[width=200pt]{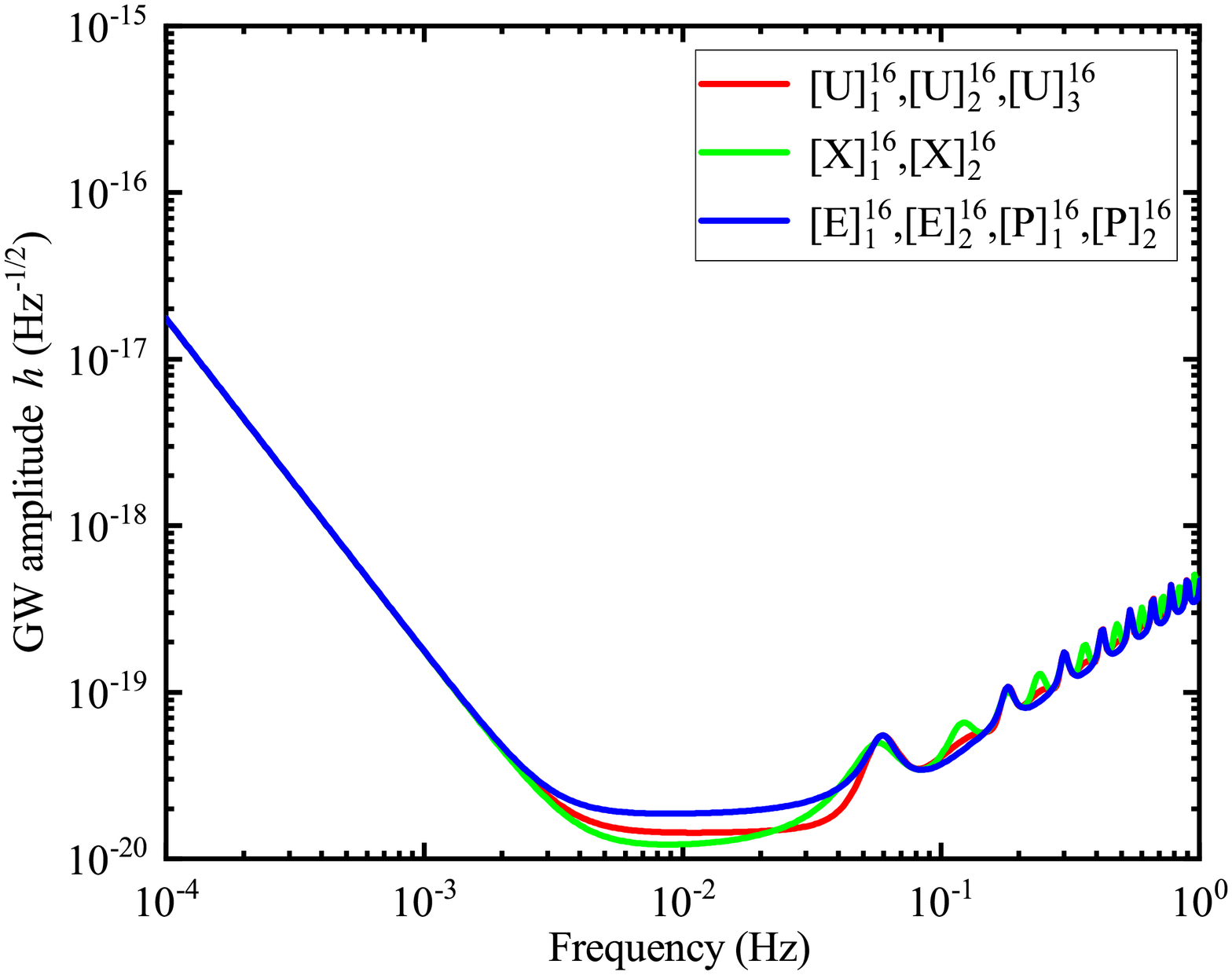}}
\end{minipage}
&
\begin{minipage}{225pt}
\centerline{\includegraphics[width=200pt]{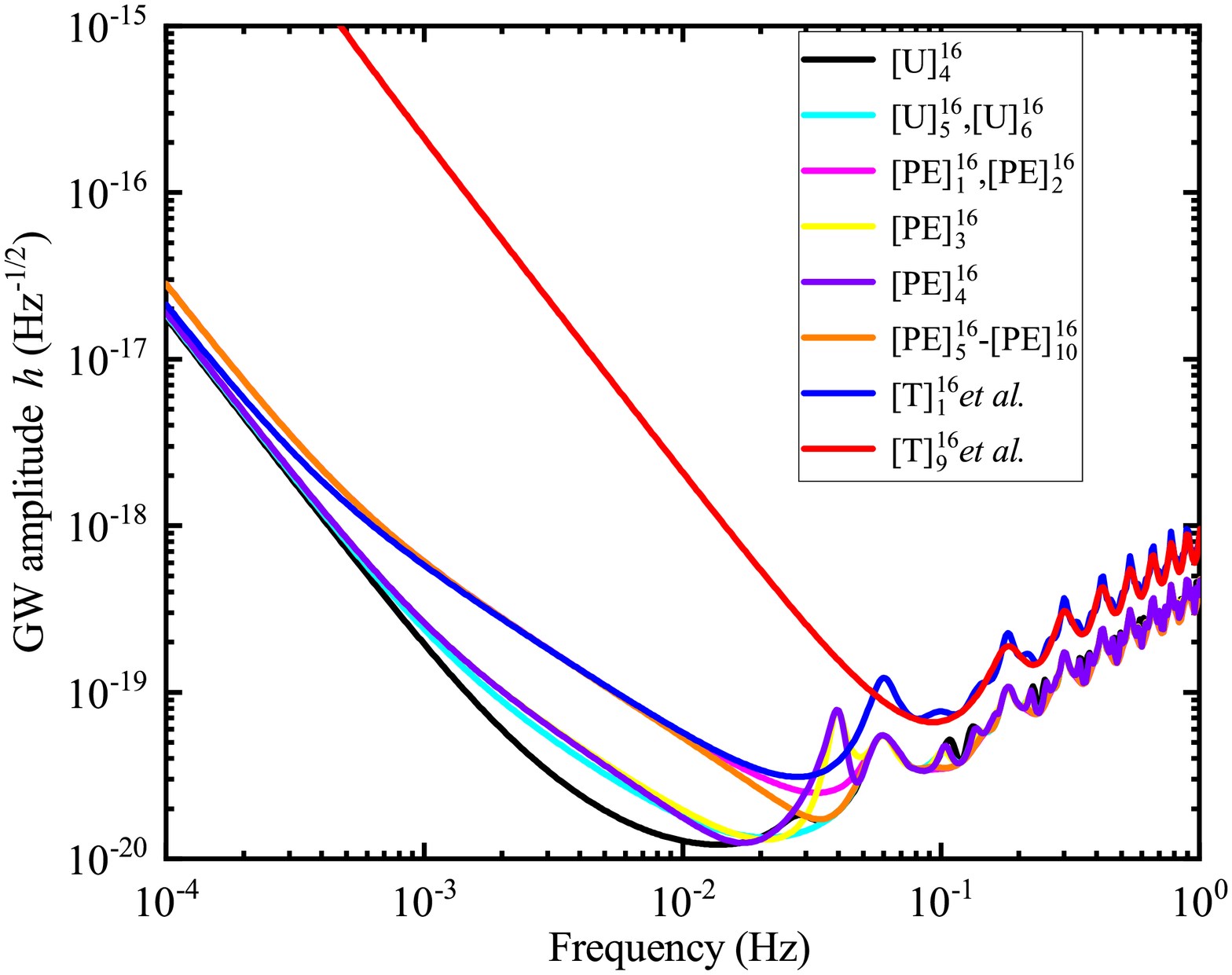}}
\end{minipage}
\end{tabular}
\renewcommand{\figurename}{Fig.}
\caption{\label{fig18}
The sensitivity curves of the sixteen-link TDI combinations.
The calculations were carried out using the parameters of the LISA detector.
The left panel shows the sensitivity curve of modified second-generation TDI combinations, while the right panel shows
that for the second-generation TDI combinations.}
\end{figure}

As given above in Eqs.~\eqref{x1x2senfun}-\eqref{EPsenfun}, there are three distinct sensitivity curves out of the nine modified second-generation TDI combinations.
The resulting curves are shown in the left plot of Fig.~\ref{fig18}.

We also analytically evaluate the GW response function and noise PSD for the second-generation TDI combinations.
Out of thirty-one TDI solutions, eight distinct sensitivity functions are obtained and enumerated as follows
\begin{align}\label{2SEN}
S(u)_1:~&[U]_4^{16},\\\notag
S(u)_2:~&[U]_5^{16}=[U]_6^{16},\\\notag
S(u)_3:~&[PE]_1^{16}=[PE]_2^{16},\\\notag
S(u)_4:~&[PE]_3^{16},\\\notag
S(u)_5:~&[PE]_4^{16},\\\notag
S(u)_6:~&[PE]_5^{16}=[PE]_6^{16}=[PE]_7^{16}=[PE]_8^{16}=[PE]_9^{16}=[PE]_{10}^{16},\\\notag
S(u)_{8}:~&[T]_1^{16}=[T]_2^{16}=[T]_3^{16}=[T]_4^{16}=[T]_5^{16}=[T]_6^{16}=[T]_7^{16}\\\notag
~&=[T]_8^{16}=[T]_{11}^{16}=[T]_{13}^{16}=[T]_{16}^{16}=[T]_{17}^{16},\\\notag
S(u)_{8}:~&[T]_9^{16}=[T]_{10}^{16}=[T]_{12}^{16}=[T]_{14}^{16}=[T]_{15}^{16}=[T]_{18}^{16}.
\end{align}
These results are shown in the right plot of Fig.~\ref{fig18}.
Judging from the sensitivity curves, overall, the modified second-generation TDI combinations are more favorable when compared to their second-generation counterparts.

\begin{figure}[!t]
\begin{tabular}{cc}
\vspace{0pt}
\begin{minipage}{225pt}
\centerline{\includegraphics[width=200pt]{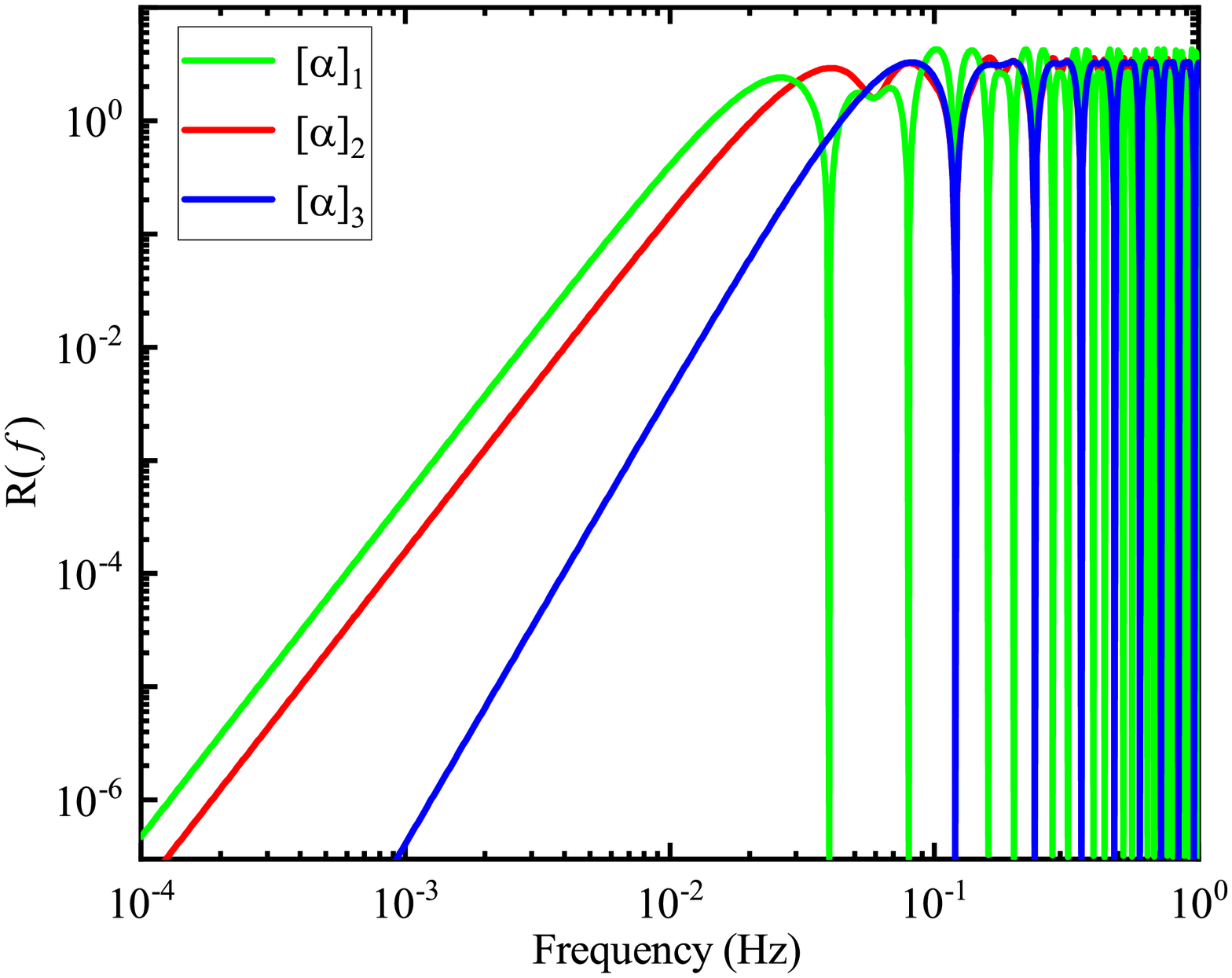}}
\end{minipage}
&
\begin{minipage}{225pt}
\centerline{\includegraphics[width=200pt]{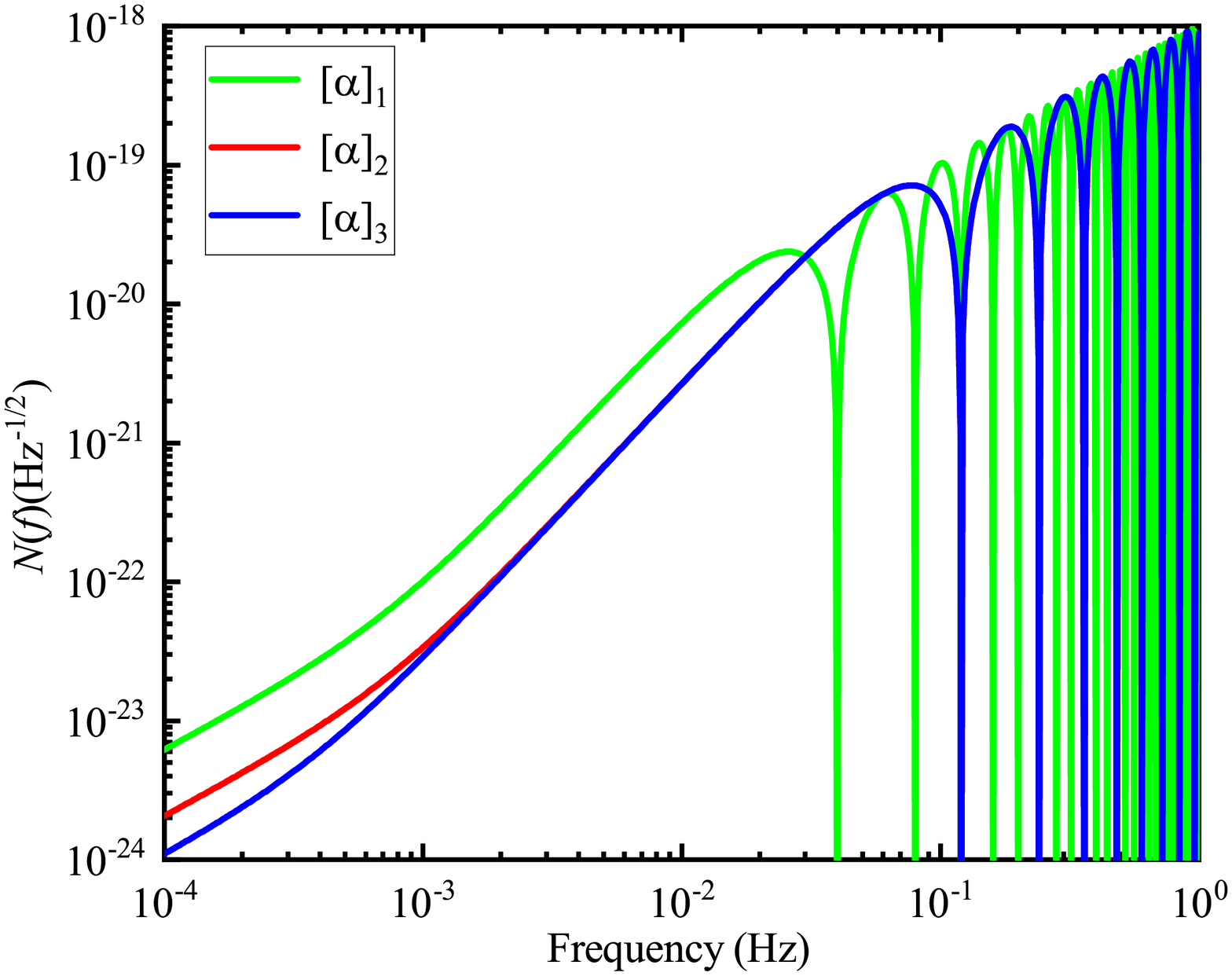}}
\end{minipage}
\end{tabular}
\renewcommand{\figurename}{Fig.}
\caption{\label{fig19}
The GW averaged response function and noise PSD of the twelve-link TDI combinations.
The calculations were carried out using the parameters of the LISA detector.
The left panel shows the  GW averaged response function, while the right panel shows
that for the noise PSD.}
\end{figure}

\begin{figure}[!t]
\includegraphics[width=0.40\textwidth]{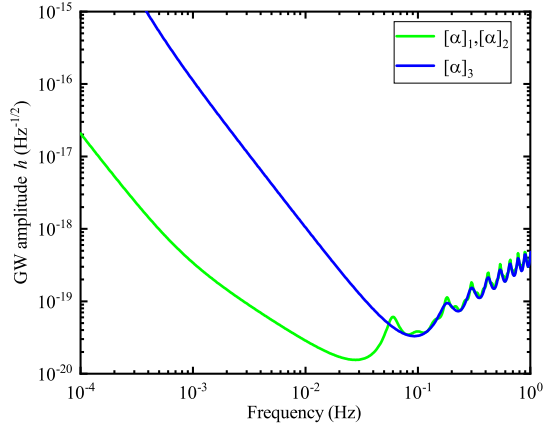}
\caption{\label{fig20}
Sensitivity curves of the twelve-link TDI combinations.
The calculations were carried out using the parameters of the LISA detector.}
\end{figure}

For the twelve-link TDI combination, the GW response and noise PSD are shown in Fig.~\ref{fig19}. 
For the standard TDI $[\alpha]_1$, the zeros are located at $nc/3L$.
On the other hand, the zeros of the alternative form $[\alpha]_2,[\alpha]_3$ are more sparsely distributed, located at $nc/L$.
Also, from Fig.~\ref{fig20}, it is observed that the alternative form $[\alpha]_2$ has the same sensitivity curve as the standard Sagnac combination $[\alpha]_1$, which outperforms $[\alpha]_3$. 

\begin{figure}[!t]
\includegraphics[width=0.40\textwidth]{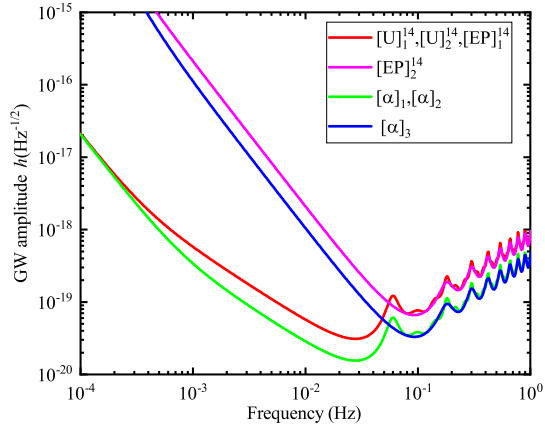}
\caption{\label{fig21}
Sensitivity curves of the fourteen-link TDI combinations and Sagnac combination $[\alpha]_1$.
The calculations were carried out using the parameters of the LISA detector.}
\end{figure}

The sensitivity curves of the fourteen-link TDI combinations and the classic twelve-link Sagnac combination are presented in Fig.~\ref{fig21}.
For the fourteen-link TDI combinations, the sensitivity functions of $[U]_1^{14},[U]_2^{14},[EP]_1^{14}$ are found to be identical, which performs better than $[EP]_2^{14}$.
However, it is found that the sensitivity functions of the fourteen-link TDI combinations are generally worse than that of the Sagnac combination.


\section{Concluding remarks}\label{section6}

By employing a ternary search implemented for the geometric TDI algorithm, this paper studies the twelve-, fourteen, and sixteen-link TDI combinations.
The obtained geometric TDI solutions, specifically the modified second-generation ones, are analyzed.
First, the laser frequency noise suppression is investigated, and the time-mismatch residual is evaluated analytically.
In particular, for the LISA detector, the derived analytic expressions are shown to be consistent with the numerical simulation results.
The modified second-generation TDI solutions investigated in the present study distinguish the cyclic directions regarding the rate of change of a given arm length.
Second, the noise PSD and GW signal response functions of relevant TDI combinations are evaluated, and the relevant TDI combinations' sensitivity functions are analytically derived.
Overall, it is observed that the sensitivity functions of the modified second-generation TDI combinations are better than the second-generation ones.

\section*{Acknowledgements}
This work is supported by the National Natural Science Foundation of China (Grant No. 11925503), Guangdong Major project of Basic and Applied Basic Research (Grant No.2019B030302001), and the Fundamental Research Funds for the Central Universities, HUST: 2172019kfyRCPY029.
We also acknowledge the financial support from Brazilian agencies 
Funda\c{c}\~ao de Amparo \`a Pesquisa do Estado de S\~ao Paulo (FAPESP),
Funda\c{c}\~ao de Amparo \`a Pesquisa do Estado do Rio de Janeiro (FAPERJ),
Conselho Nacional de Desenvolvimento Cient\'{\i}fico e Tecnol\'ogico (CNPq),
Coordena\c{c}\~ao de Aperfei\c{c}oamento de Pessoal de N\'ivel Superior (CAPES).

\appendix
\section{Index conventions}\label{apptidal}

A variety of notations for the signal channel have been utilized in the literature.
When Tinto {\it et al.} first proposed TDI~\cite{tdi-02}, a single index indicates a vertex or an arm sitting on the opposite side of the vertice.
A signal transmitted along the optical path associated with the arm $L_i$ and eventually received by the spacecraft $j$ is represented by a pair of indices ``$ij$''.
For (artificially) delayed data streams, one further introduces the notation $s_{ij,k}\equiv s_{ij}(t-L_k)$.
To be specific, $s_{31}$ gives the one-way phase difference measured at spacecraft 1, which is emitted by spacecraft 2, and subsequently transmitted along the arm $L_3$.
Such a convention has been subsequently utilized in Refs.~\cite{tdi-d22,tdi-d33,tdi-d44,tdi-d66}.
As a comparison, a second convention~\cite{tdi-d88,tdi-d99,tdi-d1010,tdi-d1111,tdi-laser-01,gwdirection-01} also makes use of two indices while adopting more intuitive notations.
Here, $s_{ij}$ denotes the phase difference measured at spacecraft $j$, which comes from spacecraft $i$.
Moreover, a third convention is also extensively utilized in the literature~\cite{tdi-clock1,tdi-clock2,tdi-clock5,tdi-filter-s4,tdi-03,tdi-clock4}.
Here, an index is used to denote the optical path sitting on the opposite side of the vertice, in the counter-clockwise direction, while an index with a prime indicates that in the clockwise direction.
Apart from the above three conventions, other notations have also been employed by some authors~\cite{tdi-laser-06}.

Obviously, the second set of conventions is rather intuitive and not restricted to the case of the three-spacecraft constellation, and the only drawback is its apparent redundancy.
On the other hand, the third choice suffices for the present purpose and is rather convenient in practice due to its compact form.
Therefore, in our manuscript, we adopt the third convention, which has also been employed in the recently updated~\cite{tdi-03}.\\

\section{The delay operator polynomial coefficients for the modified second-generation TDI combinations}\label{apptida2}
This Appendix presents the Fourier transforms of the delay operator polynomial coefficients for the modified second-generation TDI combinations.
The corresponding coefficients ${\tilde P_{i}(u)}$, ${\tilde P_{i'}(u)}$ of $[X]_1^{16},[X]_2^{16}$ combinations are, respectively,
\begin{align}\label{x1coff}
\tilde P_1(u)=&1 - e^{2iu} -  e^{4iu} + e^{6iu},\\\notag
\tilde P_2(u)=&0,\\\notag
\tilde P_3(u)=& -e^{iu} + e^{3iu} + e^{5iu} -e^{7iu},\\\notag
\tilde P_{1'}(u)=& - 1 +  e^{2iu} +  e^{4iu} - e^{6iu},\\\notag
\tilde P_{2'}(u)=& e^{iu} -  e^{3iu} -  e^{5iu} + e^{7iu},\\\notag
\tilde P_{3'}(u)=&0,
\end{align}
and
\begin{align}\label{x2coff}
\tilde P_1(u) =& 1 - 2 e^{2iu} +  e^{4iu},\\\notag
\tilde P_2(u) =& 0,\\\notag
\tilde P_3(u) =&  -  e^{iu} + 2 e^{3iu} -  e^{5iu},\\\notag
\tilde P_{1'}(u) =&  - 1 + 2e^{2iu} - e^{4iu},\\\notag
\tilde P_{2'}(u)=& e^{iu} - 2 e^{3iu} +  e^{5iu},\\\notag
\tilde P_{3'}(u) =& 0.
\end{align}
The corresponding coefficients ${\tilde P_{i}(u)}$, ${\tilde P_{i'}(u)}$ of $[U]_1^{16},[U]_2^{16},[U]_3^{16}$ combinations are, respectively,
\begin{align}\label{u1coff}
\tilde P_1(u) =& 1 - 2e^{3iu} + e^{6iu},\\\notag
\tilde P_2(u) =& 0,\\\notag
\tilde P_3(u) =& 0,\\\notag
\tilde P_{1'}(u)  =&  - 1 + e^{2iu} + e^{3iu} -e^{5iu},\\\notag
\tilde P_{2'}(u) =& e^{iu} - e^{2iu}-e^{4iu} + e^{5iu},\\\notag
\tilde P_{3'}(u) =&  -e^{iu} + e^{3iu} + e^{4iu} - e^{6iu},
\end{align}

\begin{align}\label{u2coff}
\tilde P_1(u) =& 1 - e^{2iu}- e^{3iu}+ e^{5iu},\\\notag
\tilde P_2(u) =& 0\\\notag
\tilde P_3(u) =& 0\\\notag
\tilde P_{1'}(u) =&  - 1 + 2e^{2iu} - e^{4iu}\\\notag
\tilde P_{2'}(u)=& e^{iu} - e^{2iu} - e^{3iu} + e^{4iu}\\\notag
\tilde P_{3'}(u)=&- e^{iu} + 2e^{3iu} - e^{5iu},
\end{align}
and
\begin{align}\label{u3coff}
\tilde P_1(u) =& 1 - e^{iu} - e^{3iu}+ e^{4iu},\\\notag
\tilde P_2(u) =& 0,\\\notag
\tilde P_3(u)=& 0,\\\notag
\tilde P_{1'}(u) =&  - 1 + e^{iu} + e^{2iu} - e^{3iu},\\\notag
\tilde P_{2'}(u) =& e^{iu}- 2e^{2iu} + e^{3iu},\\\notag
\tilde P_{3'}(u)=&  - e^{iu} + e^{2iu} + e^{3iu}- e^{4iu}.
\end{align}

The corresponding coefficients ${\tilde P_{i}(u)}$, ${\tilde P_{i'}(u)}$ of $[E]_1^{16},[E]_2^{16}$  combinations are, respectively,
\begin{align}\label{e1coff}
\tilde P_1(u) =& 1 - 2e^{2iu} + e^{4iu},\\\notag
\tilde P_2(u)=& e^{iu}- e^{2iu}- e^{3iu} + e^{4iu},\\\notag
\tilde P_3(u) =& 0,\\\notag
\tilde P_{1'}(u)=&  - 1 + 2e^{2iu} - e^{4iu},\\\notag
\tilde P_{2'}(u) =& 0,\\\notag
\tilde P_{3'}(u) =&  - e^{iu} + e^{2iu}+ e^{3iu} - e^{4iu},
\end{align}
and
\begin{align}\label{e2coff}
\tilde P_1(u) =& 1 - e^{iu} - e^{2iu} + e^{3iu},\\\notag
\tilde P_2(u) =& e^{iu} - 2e^{2iu} + e^{3iu},\\\notag
\tilde P_3(u) =& 0,\\\notag
\tilde P_{1'}(u)=&  - 1 + e^{iu} + e^{2iu} - e^{3iu},\\\notag
\tilde P_{2'}(u) =& 0,\\\notag
\tilde P_{3'}(u)=&  - e^{iu} + 2e^{2iu} - e^{3iu}.
\end{align}
The corresponding coefficients ${\tilde P_{i}(u)}$, ${\tilde P_{i'}(u)}$ of $[P]_1^{16},[P]_2^{16}$ combinations are, respectively,
\begin{align}\label{p1coff}
\tilde P_1(u) =& 1 - e^{iu} - e^{2iu} + e^{3iu},\\\notag
\tilde P_2(u) =& 1 - 2e^{2iu}+ e^{4iu},\\\notag
\tilde P_3(u) =& 0,\\\notag
\tilde P_{1'}(u)=&  - 1 + 2e^{2iu} - e^{4iu},\\\notag
\tilde P_{2'}(u) =&  - 1 + e^{iu}+ e^{2iu} - e^{3iu},\\\notag
\tilde P_{3'}(u) =& 0,
\end{align}
and
\begin{align}\label{p2coff}
\tilde P_1(u)=& 1 - 2e^{iu} + e^{2iu},\\\notag
\tilde P_2(u) =& 1 - {e^{iu}} - e^{2iu} + e^{3iu},\\\notag
\tilde P_3(u)=& 0\\\notag
\tilde P_{1'}(u) =& - 1 + e^{iu} + e^{2iu}- e^{3iu},\\\notag
\tilde P_{2'}(u) =&  - 1 + 2e^{iu} - e^{2iu},\\\notag
\tilde P_{3'}(u) =& 0.
\end{align}

\bibliographystyle{h-physrev}
\bibliography{reference_wang}
\end{document}